\documentclass[
,showpacs ,amssymb,superscriptaddress,aps
]{revtex4}
\usepackage{graphicx}
\usepackage{color}
\input{epsf}

\usepackage{amsmath,amssymb}
\usepackage{bm}
\usepackage{times}
\usepackage{cancel}
\newcommand{\dalm}{\kern1pt\vbox{\hrule height 0.9pt\hbox{\vrule width 0.9pt
\hskip 2.5pt\vbox{\vskip 5.5pt}\hskip 3pt\vrule width 0.3pt}\hrule height 0.3pt}
\kern1pt}

\begin{document}



\title{
Dependence of outer boundary condition on protoneutron star asteroseismology with gravitational-wave signatures
}

\author{Hajime Sotani}
\email{hajime.sotani@nao.ac.jp}
\affiliation{Division of Theoretical Astronomy, National Astronomical Observatory of Japan, 2-21-1 Osawa, Mitaka, Tokyo 181-8588, Japan}

\author{Takami Kuroda}
\affiliation{Institut f\"ur Kernphysik, Technische Universit\"at Darmstadt, Schlossgartenstrasse 9, 64289 Darmstadt, Germany}

\author{Tomoya Takiwaki}
\affiliation{Division of Theoretical Astronomy, National Astronomical Observatory of Japan, 2-21-1 Osawa, Mitaka, Tokyo 181-8588, Japan}
\affiliation{Center for Computational Astrophysics, National Astronomical Observatory of Japan, 2-21-1 Osawa, Mitaka, Tokyo 181-8588, Japan}

\author{Kei Kotake}
\affiliation{Department of Applied Physics, Fukuoka University, 8-19-1, Jonan, Nanakuma, Fukuoka, 814-0180, Japan}

\date{\today}

\begin{abstract}
To obtain the eigenfrequencies of a protoneutron star (PNS) in the postbounce phase of  core-collapse supernovae (CCSNe), we perform a linear perturbation analysis of the angle-averaged PNS profiles using results from a general relativistic CCSN simulation of a $15 M_{\odot}$ star.  In this work, we investigate how the choice of the outer boundary condition could affect the PNS oscillation modes in the linear analysis. By changing the density at the outer boundary of the PNS surface in a parametric manner, we show that the eigenfrequencies strongly depend on the surface density. By comparing with the gravitational wave (GW) signatures obtained in the hydrodynamics simulation,  the so-called surface $g$-mode of the PNS can be well ascribed to the fundamental oscillations of the PNS. The frequency of the fundamental oscillations can be fitted by a function of the mass and radius of the PNS similar to the case of cold neutron stars. In the case that the position of the outer boundary is chosen to cover not only the PNS but also the surrounding postshock region,  we obtain the eigenfrequencies close to the modulation frequencies of the standing accretion-shock instability (SASI). However, we point out that these oscillation modes are unlikely to have the same physical origin of the SASI modes seen in the hydrodynamics simulation. We discuss possible limitations of applying the angle-averaged, linear perturbation analysis to extract the full ingredients of the CCSN GW signatures.
\end{abstract}

\pacs{04.40.Dg, 97.10.Sj, 04.30.-w}
%
\maketitle
\section{Introduction}
\label{sec:I}

Success of direct observations of gravitational waves (GWs) from the compact binary mergers ushered in a new era of GW astronomy. Up to now, GWs from five binary black hole (BH) mergers, i.e., GW150914 \cite{GW1}, GW151226 \cite{GW2}, GW170104 \cite{GW3}, GW170608 \cite{GW4}, and GW170814 \cite{GW5}, and one binary neutron star (NS) merger, i.e., GW170817 \cite{GW6}, have been detected by LIGO (Laser Interferometer Gravitational-wave Observatory) Scientific Collaboration and Virgo Collaboration. In the event of GW170817 \cite{EM}, the electromagnetic-wave counterpart has been detected, which opens yet another new era of multi-messenger astronomy. 
In addition to the advanced LIGO and advanced Virgo, KAGRA will be operational in the coming years \cite{aso13}. Furthermore, the third-generation detectors have been proposed  such as Einstein Telescope and Cosmic  Explorer~\cite{punturo,CE}. At such high level of precision, these detectors are sensitive enough to a wide variety of compact objects. Next to the primary targets of the compact binary coalescence, other intriguing sources include core-collapse supernovae (CCSNe) \cite{KotakeGWreview}, which mark the catastrophic end of massive stars and produce all these compact objects.

In order to study the GW signatures from CCSNe, numerical simulations have been done extensively (e.g., \cite{MJM2013,CDAF2013,KKT2016,Ott13,Andresen16,Murphy09,Yakunin15,OC2018}). The most distinct GW emission process commonly seen in recent self-consistent three-dimensional (3D) models is associated with the excitation of core/protoneutron star (PNS) oscillatory modes \cite{KKT2016,radice2018,Andresen18}. This is supported by the evidence that the Brunt-V\"{a}is\"{a}l\"{a} frequency estimated at the PNS surface is in good accordance with that of the strongest GW component. The typical GW frequency of the surface $g$-mode is approximately expressed by $GM_{\rm PNS}/R^2_{\rm PNS}$~\cite{MJM2013,CDAF2013,KKT2016,Murphy09} with $G$ the gravitational constant, $M_{\rm PNS}$ and $R_{\rm PNS}$ the mass and radius of the PNS, respectively. In the postbounce phase, the PNS mass increases with time due to the mass accretion and the PNS radius decreases with time due to the mass accretion onto the PNS and neutrino cooling. Accordingly, the typical GW frequency of the surface $g$-mode increases with time after bounce \cite{MJM2013,Murphy09}, which is roughly in the range of $\sim 500-1000$ Hz. These oscillations are excited because the PNS surface is chimed by the mass motions. Recent studies indicate that the dominant excitation  process may be sensitive to the spacial dimension in the hydrodynamics simulations. In axisymmetric two-dimensional (2D) models, the mass accretion from above the PNS, where the mass accretion activity to the PNS is influenced by the growth of neutrino-driven convection and the standing accretion-shock instability (SASI)~\cite{Blondin03,Foglizzo06}, are the main excitation process of the PNS surface oscillations \cite{Murphy09,MJM2013}. While Ref.~\cite{Andresen16} showed in the 3D models that the PNS convection could also significantly contribute to the postbounce GW emission. In addition to the PNS oscillations, recent 3D CCSN models have shown another remarkable GW signatures whose frequencies are close to the modulation frequency of the SASI motion, i.e., $\sim 100$ Hz~\cite{KKT2016,Andresen16,OC2018,radice2018}.  Thus, the detection of the GWs with $\sim 100$ Hz separately from those with $\sim 500-1000$ Hz may provide a probe into the SASI activity in the pre-explosion supernova core \citep{KKT2016,Andresen16}.

The hydrodynamics modeling is really powerful to clarify the inner-workings of the forming compact objects, while linear perturbation approaches are also valuable to understand the physics behind the numerical results obtained by simulations. Given angle-averaged profiles obtained in hydrodynamics models, oscillation spectra are determined by a linear analysis. Then, if one could find a correlation between the properties of the background model and the resultant oscillation spectra, one can extract the information about the background model through observations of the spectra. This technique is known as asteroseismology, which has been extensively investigated in the context of cold NSs. 
With this technique, it has been suggested that the properties of the NSs such as the mass ($M$), radius ($R$), and EOS, would be constrained with GW asteroseismology, where one would get an information about the source object with the GW spectra (e.g., \cite{AK1996,AK1998,KS1999,STM2001,SH2003,SYMT2011,PA2012,DGKK2013}).

Compared to a lot of studies with the linear perturbation analysis on cold NSs, similar studies on PNSs are very few \cite{FMP2003,FKAO2015,ST2016,Camelio17,SKTK2017,TCPF2018,MRBV2018,TCPOF2019}. The paucity of the perturbative studies on the PNSs may come from the difficulty for preparing for the background model of the PNSs. That is, unlike the case of nearly hydrostatic cold NSs, one also needs the time dependent radial distributions of the electron fraction and, e.g., the entropy per baryon for constructing the finite temperature PNS models. However, these time dependent spatial profiles are determined only via the self-consistent CCSN simulations which are computationally expensive. 
Among the recent studies to tackle with this problem \cite{FMP2003,FKAO2015,ST2016,Camelio17,SKTK2017,TCPF2018,MRBV2018,TCPOF2019}, we have found that the frequencies of the fundamental $(f)$ and the space-time $(w)$ modes \cite{ST2016,SKTK2017} can be respectively expressed as a function of the average density and compactness of the PNSs almost independently of the EOS of PNSs, in a similar way to the case of cold NSs \cite{AK1996,AK1998}.  In this context, a universal relation of the CCSN GW spectra is recently reported in Ref. \cite{TCOMF2019a}.

Up to now, two representative ways have been proposed for constructing the background PNS models for determining the eigenfrequencies in the linear perturbation analysis. They differ in the definition of the PNS surface. One is the PNS model, in which the surface density is fixed as a specific value, for example, of $\sim10^{10}$ g cm$^{-3}$ \cite{ST2016,SKTK2017,MRBV2018}. In this case, one can impose the boundary condition similarly as taken in the stellar oscillation analysis and can classify the stellar oscillations. However, unlike the usual cold NS case, the low density matter still hovers and the accretion shock also exists outside this density region, whose influences might not be negligible. Therefore the PNS model covering up to the shock radius is also proposed \cite{TCPF2018,TCPOF2019}. With the boundary condition imposed at the shock, one can investigate the global oscillations inside the whole postshock region, although the eigenvalue problem to solve is significantly different from that with the PNS model with the fixed surface density.

In this study, we calculate the eigenfrequencies in the PNS models by the linear perturbation analysis with the two different boundary conditions, i.e., either at the PNS surface with a fixed specific density or at the shock radius, with an attempt to identify the excitation mechanism of the GW signatures seen in the numerical simulation. This paper is organized as follows. Section \ref{PNSmodel} starts with a brief summary of the PNS models employed in this work. In Section \ref{sec:III}, we describe the linear perturbation analysis to solve the eigenvalue problem. Section \ref{sec:IV} presents our results and the comparison with the GW signal computed in the numerical simulations. We summarize our results and discuss their implications in Section \ref{sec:V}. Unless otherwise mentioned, we adopt geometric units in the following, $c=G=1$, where $c$ denotes the speed of light, and the metric signature is $(-,+,+,+)$. The time is measured after bounce 
($T_{\rm pb} = 0)$

\section{PNS Models}
 \label{PNSmodel}
The line element is expressed as
\begin{equation}
  ds^2=-\alpha^2dt^2+\gamma_{ij}(dx^i+\beta^idt)(dx^j+\beta^jdt),  \label{eq:ds_BSSN}
\end{equation}
where $\alpha$, $\beta^i$, and $\gamma_{ij}$ are the lapse, shift vector, and three metric, respectively. To prepare the background of PNS models, the metric functions $\alpha$, $\beta^i$, and $\gamma_{ij}$ from hydrodynamics simulations, which are not spherically symmetric, are transformed into the spherically symmetric properties, assuming that the hydrodynamic background at each time step is also static and spherically symmetric. In this procedure, all variables defined on the Cartesian coordinates
 in numerical relativity simulations are transformed into those in polar coordinates by spatially linear interpolation
at each time step. Then, the space-time in the isotropic coordinates can be rewritten as
\begin{align}
  ds^2 =&-\alpha^2 dt^2+ \gamma_{\hat{r}\hat{r}} 
  (d\hat{r}^2+\hat{r}^2d\theta^2+\hat{r}^2\rm{sin}^2\theta d\phi^2),
  \label{eq:ds_isotropic}
\end{align}
where $\hat{r}$ denotes the isotropic radius $\hat{r}=\sqrt{x^2+y^2+z^2}$.


In the calculations of stellar oscillations, we adopt the following spherically symmetric space-time
\begin{equation}
  ds^2 =-e^{2\Phi} dt^2 + e^{2\Lambda} dr^2 + r^2\left(d\theta^2 + \sin^2\theta d\phi^2\right), \label{eq:ds_spherical}
\end{equation}
as a background space-time, where $\Phi$ and $\Lambda$ are functions of only $r$. We remark that the metric expressed by Eq. (\ref{eq:ds_spherical}) is similar to the Schwarzschild metric and is given by the coordinate transformation from the isotropic coordinates, i.e. Eqs. (\ref{eq:ds_BSSN}) or (\ref{eq:ds_isotropic}).
Additionally, the metric function $\Lambda$ is associated with the mass function $m$ in such a way that $e^{-2\Lambda}=1-2m/r$. Then, the background four-velocity of the fluid element is given by $u^\mu=(e^{-\Phi},0,0,0)$.
Comparing  Eqs.(\ref{eq:ds_isotropic}) and (\ref{eq:ds_spherical}), the conversion relation is expressed as followings
\begin{flalign}
& &&e^{2\Phi} =\alpha^2,&& \label{eq:conv_1}\\
&  &&r^2=\gamma_{\hat{r}\hat{r}}\hat{r}^2,&& \label{eq:conv_2}\\
&\text{and}&&&&  \nonumber \\
&  &&e^{2\Lambda}dr^2=\gamma_{\hat{r}\hat{r}}d\hat{r}^2.&&  \label{eq:conv_3}
  \end{flalign}
From these, one can deduce the following relations
\begin{flalign}
dr=&\left(\gamma_{\hat{r}\hat{r}}+\frac{\hat{r}}{2}\frac{\partial \gamma_{\hat{r}\hat{r}}}{\partial \hat{r}}\right)\frac{\hat{r}}{r}d\hat{r},\label{eq:dr}\\
    m=&\left[1-\frac{(\gamma_{\hat{r}\hat{r}}+\hat{r}\partial_{\hat{r}}\gamma_{\hat{r}\hat{r}}/2)^2}{\gamma_{\hat{r}\hat{r}}^2}\right] \frac{\gamma_{\hat{r}\hat{r}}^{1/2}}{2}\hat{r}. \label{eq:m}
\end{flalign}
In this study, instead of using Eq. (\ref{eq:m}), we evaluate the enclosed gravitational mass $m$
 within $\hat r$ and use a simple conversion relation $r=\hat r(1+m/2\hat r)^2$, from isotropic to Schwarzschild coordinates.
 Although this simple conversion relation can originally be applied to the exterior of the object, we employ it as it can suppress the high frequency structural noise that appears when using Eq. (\ref{eq:m}) without some appropriate smoothing. Since we use the spatial derivative of $\Lambda$ that is a function of $m$ in the following seismology analysis, spurious noise should be suppressed. We consider that the difference between the correct, i.e. Eq. (\ref{eq:m}), and simple evaluations is not so significant. The highest values of $\exp{(2\Lambda)}=(1-2m/r)^{-1}$ appear at $\hat r\sim 1.3\times10^6$ cm and they differ approximately 1 \% between both evaluations.

\begin{figure*}[tbp]
\begin{center}
\begin{tabular}{ccc}
\includegraphics[scale=0.43]{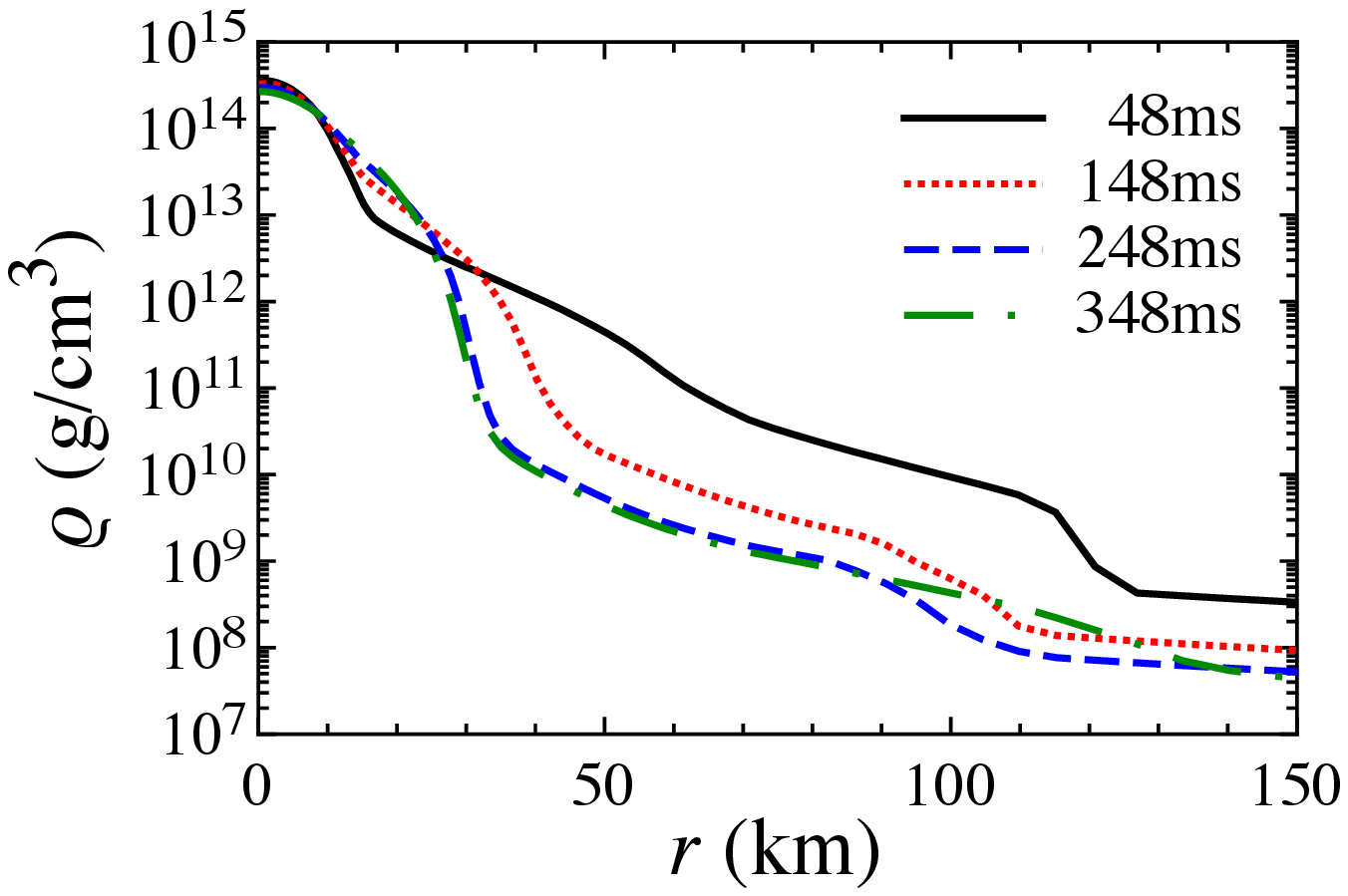} &
\includegraphics[scale=0.43]{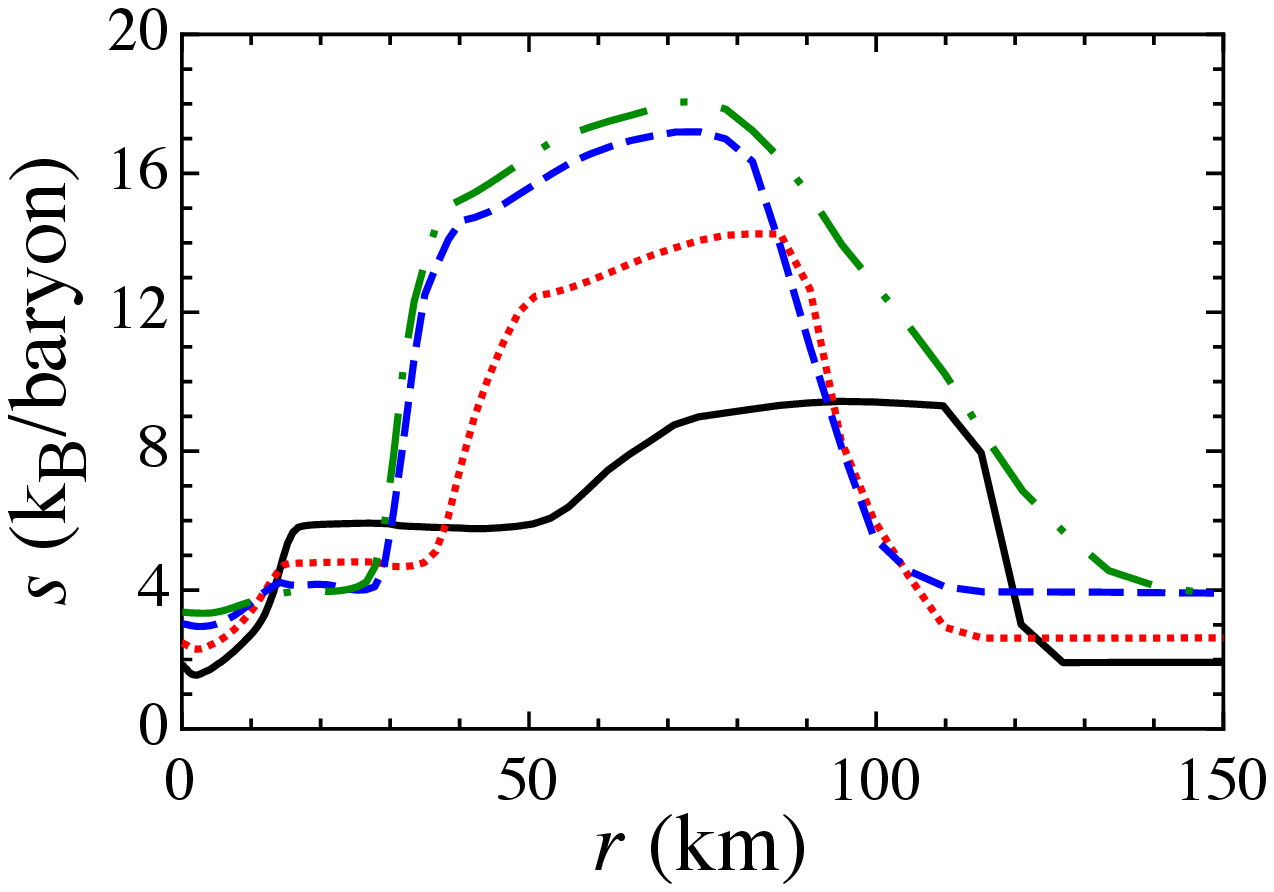} &
\includegraphics[scale=0.43]{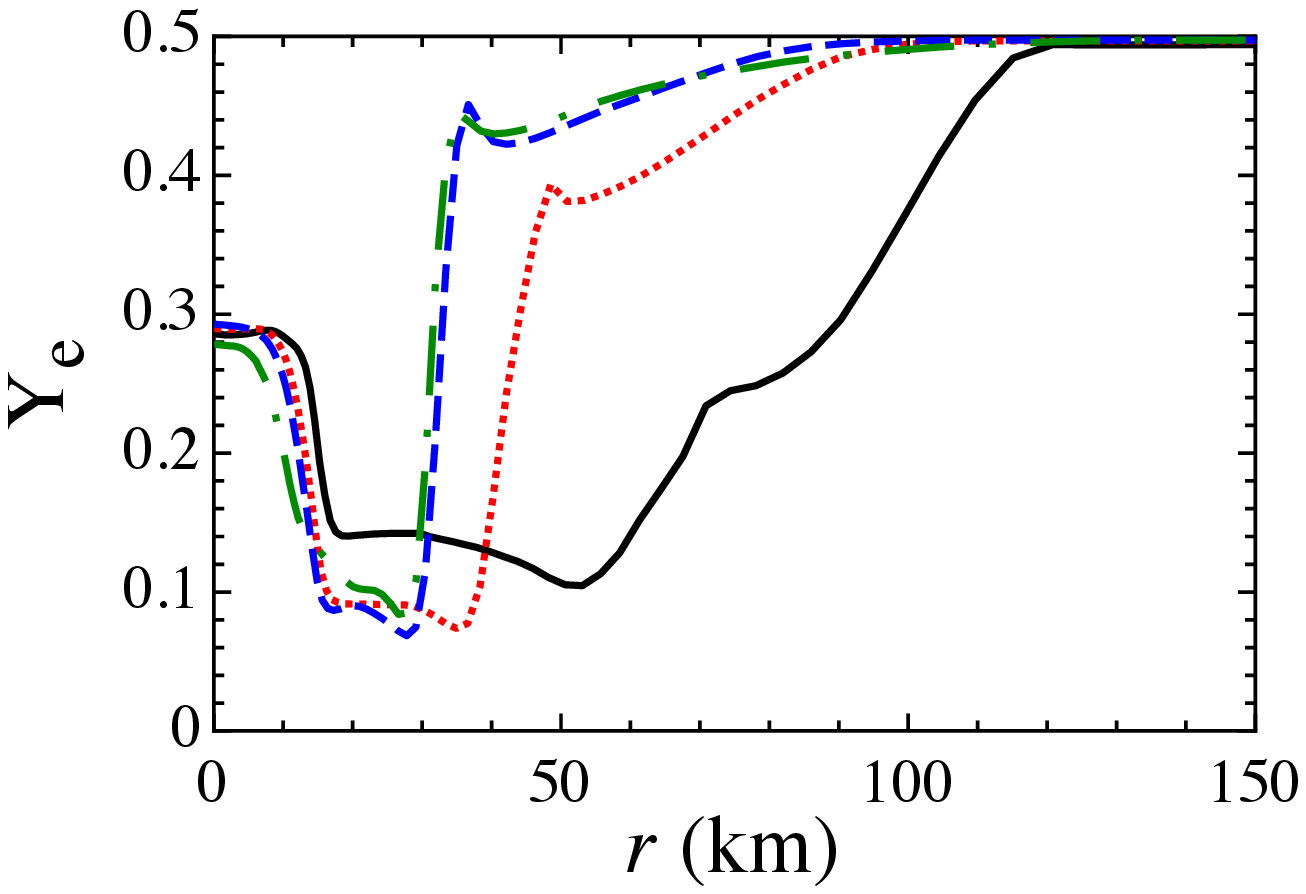}  
\end{tabular}
\end{center}
\caption{
(Spherically-averaged) radial profiles of the rest mass density ($\rho$), entropy per baryon ($s$), and electron faction ($Y_e$) at 48, 148, 248, and 348 ms after core bounce for a 3D-GR model of SFHx in \cite{KKT2016}.
}
\label{fig:back}
\end{figure*}

\begin{figure*}[tbp]
\begin{center}
\begin{tabular}{cc}
\includegraphics[scale=0.55]{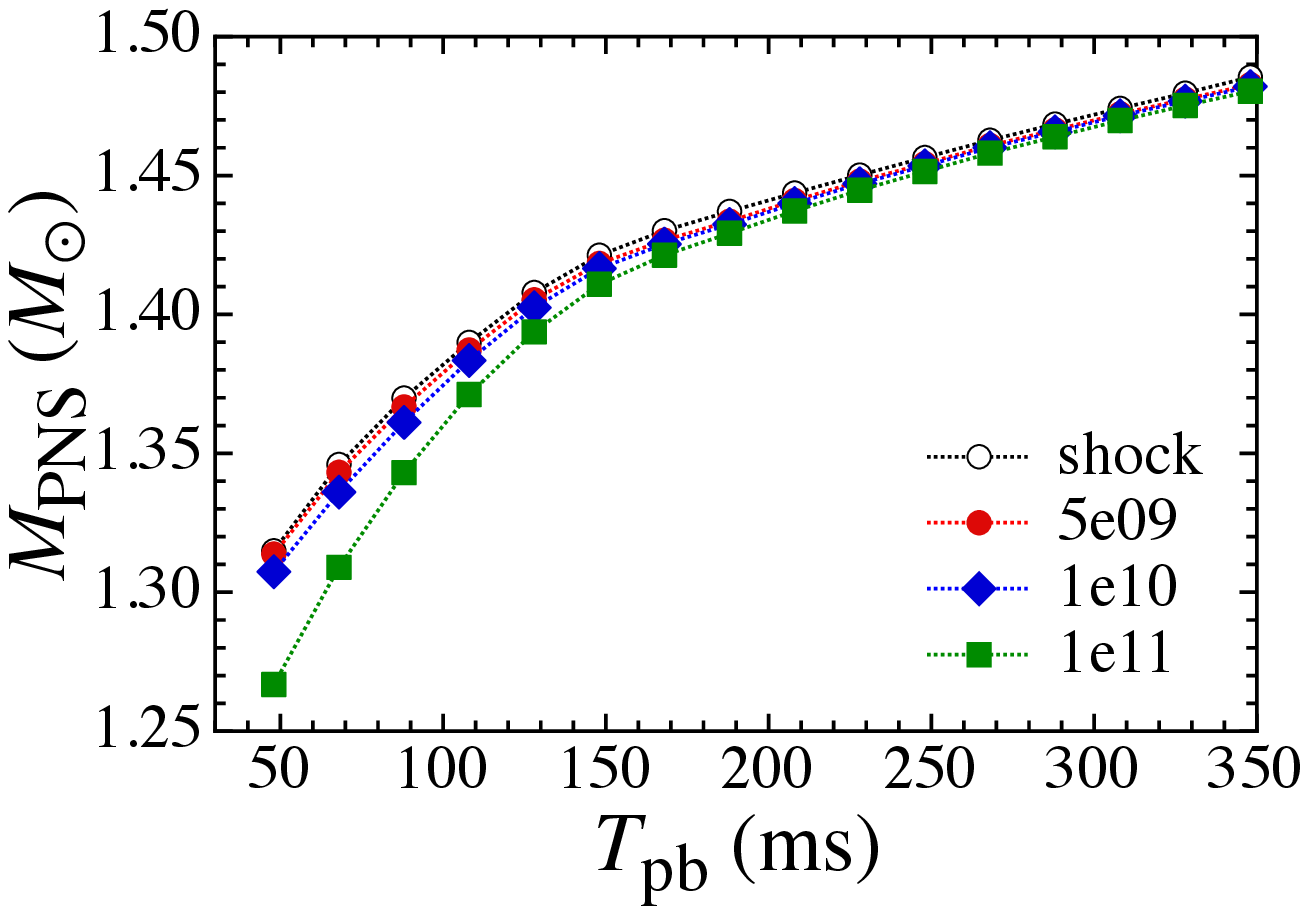} &
\includegraphics[scale=0.55]{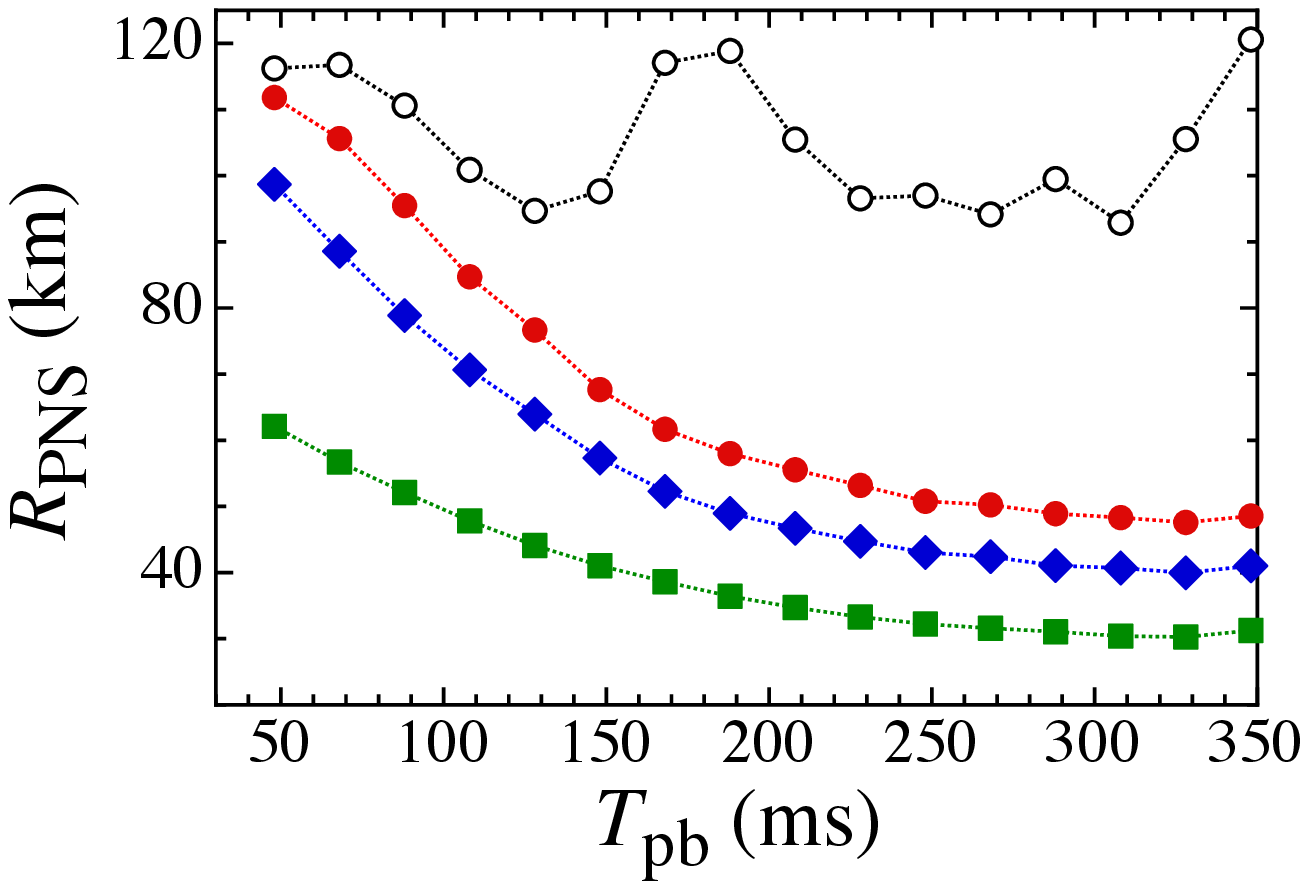}  
\end{tabular}
\end{center}
\caption{
Similar to Figure 1, but for the time evolution of the PNS gravitational mass (left panel) and radius (right panel) as a function of the postbounce time. The different lines correspond to the different definitions of the PNS model, i.e., $\rho_s=5\times 10^9$ (filled-circle), $10^{10}$ (diamond), $10^{11}$ g cm$^{-3}$ (square), and at the shock (open-circle).
}
\label{fig:MtRt}
\end{figure*}

In the present study, we especially focus on the numerical results constructed with SFHx EOS \cite{SHF2013}. The initial hydrodynamic profile is taken from a $15M_\odot$ progenitor model \cite{WW95} in the simulation \cite{KKT2016}. In Fig. \ref{fig:back}, we show the radial profiles of the rest mass density $\rho$, entropy per baryon $s$, and electron fraction $Y_e$ at several time snapshots after bounce. From this figure, one can observe that the profiles at 248 ms is almost the same as that at 348 ms. On these background properties, we consider the specific oscillations in PNS at each time step. As PNS models, we consider two different approaches, i.e., 1) as in Ref. \cite{MRBV2018}, the position, where the rest mass density is equivalent to be $\rho_s=5\times 10^9$, $10^{10}$, and $10^{11}$ g cm$^{-3}$, is considered as the stellar surface of a background PNS, or 2) the domain inside the shock radius is adopted for calculating the frequencies of stellar oscillations as in Refs. \cite{TCPF2018,TCPOF2019}. Here, we define the position of the shock radius, where the entropy per baryon becomes $s=7$ $k_{\rm B}$ baryon$^{-1}$ at the outermost radial position with excluding obviously infalling unshocked stellar mantles. In Fig. \ref{fig:MtRt}, we show the time evolution of the PNS gravitational mass and radius, which are determined with different definitions of the PNS surface, as a function of the postbounce time $T_{\rm pb}$. One can observe that the gravitational masses after $\sim 150$ ms are almost independent from the definition of PNS surface, while the PNS radius still depends on the surface density. In the right panel of Fig. \ref{fig:MtRt}, we also show the shock radius, which does not change monotonically with time due to the vigorous SASI motion \cite{KKT2016}.

\section{Perturbation equations in the Cowling approximation}
\label{sec:III}
\color{black}

In this paper, we simply assume the relativistic Cowling approximation \cite{Finn1988}, i.e., the metric perturbations are neglected during the stellar oscillations, where the oscillation frequencies can  be discussed qualitatively but the damping of oscillations (or the imaginary part of complex frequencies) due to the GW emission can not be calculated.  
We remark that our perturbation formalism is basically the same as in Ref. \cite{MRBV2018} with $\delta\hat{\alpha}=0$, noting that this should be improved in our future work as in \cite{MRBV2018}.
The Lagrangian displacement vector of fluid element $\xi^i$ for the polar type oscillations is given by
\begin{equation}
  \xi^i(t,r,\theta,\phi) = \left(e^{-\Lambda}W, -V\partial_\theta, -\frac{V}{\sin^2\theta}\partial_\phi\right)\frac{1}{r^2}Y_{\ell k}(\theta,\phi),
\end{equation}
where $W$ and $V$ are a function of $t$ and $r$, while $Y_{\ell k}(\theta,\phi)$ denotes the spherical harmonics with the azimuthal quantum number $\ell$ and the magnetic quantum number $k$. With $\xi^i$, one can obtain the perturbed four-velocity $\delta u^\mu$ as
\begin{equation}
  \delta u^\mu = \left(0, e^{-\Lambda}\dot{W}, -\dot{V}\partial_\theta, -\frac{\dot V}{\sin^2\theta}\partial_\phi\right)
      \frac{1}{r^2}e^{-\Phi}Y_{\ell k},   \label{eq:du}
\end{equation}
where the dot denotes the partial derivative with respect to $t$. In addition, one should add the perturbations of the baryon number density $n_{\rm b}$, the pressure $p$, and the energy density $\varepsilon$.

\begin{figure*}[tbp]
\begin{center}
\begin{tabular}{cc}
\includegraphics[scale=0.5]{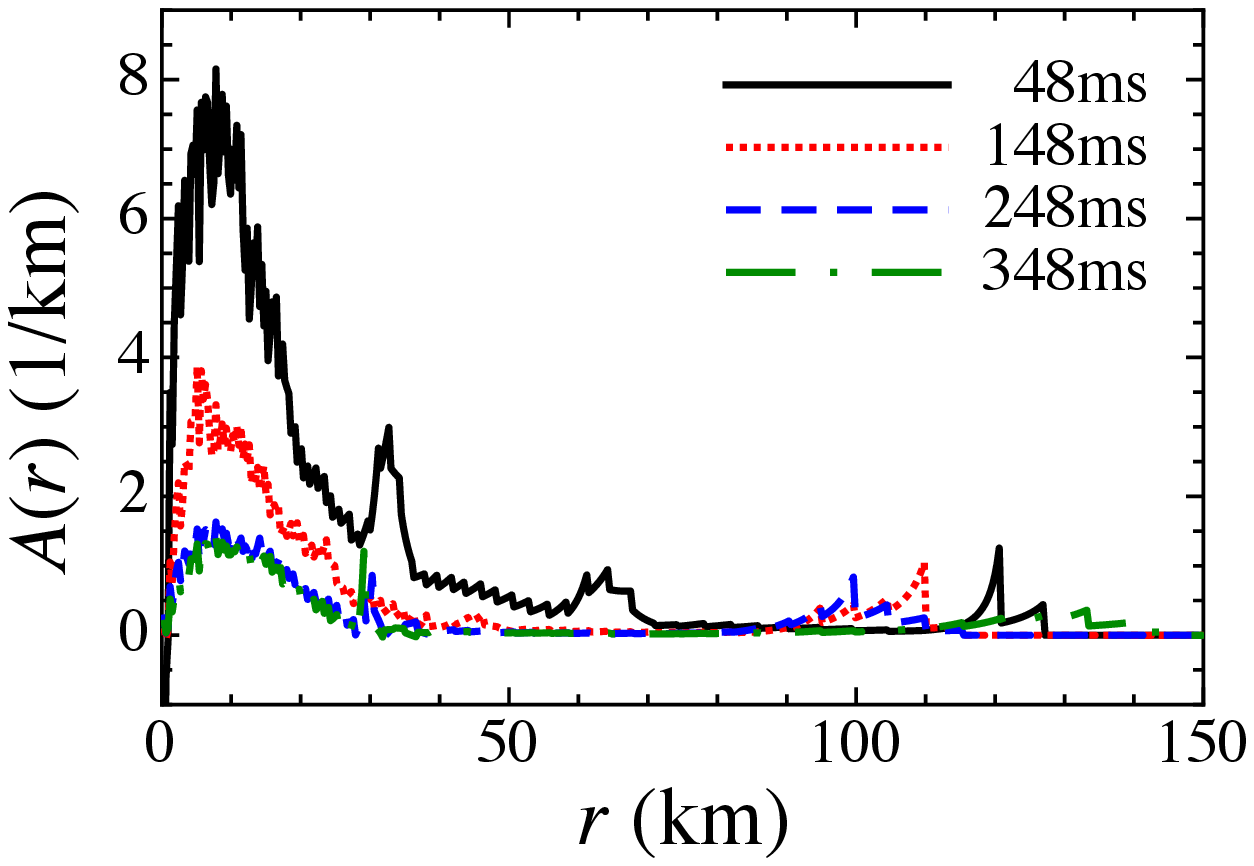} &
\includegraphics[scale=0.5]{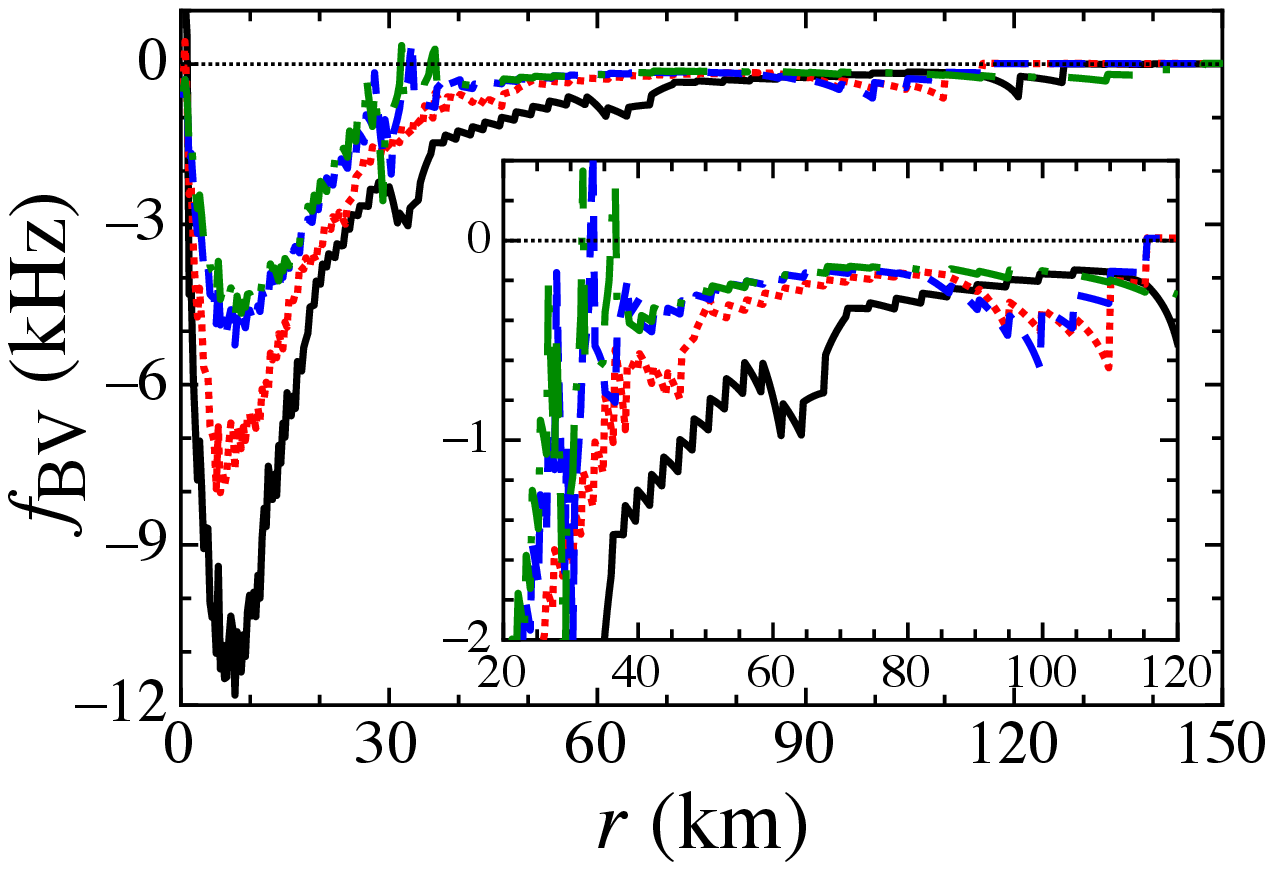}  
\end{tabular}
\end{center}
\caption{
(Spherically averaged) radial profiles of the relativistic Schwarzschild discriminant ${\cal A}$ (left panel) and Brunt-V\"{a}is\"{a}l\"{a} frequency (right panel) at $T_{\rm pb}=48$, 148, 248, and 348ms. The inset in the right panel is just a zoom-up, focusing on the frequency range below 2 kHz in the absolute value. 
}
\label{fig:BV}
\end{figure*}

From the baryon number conservation with the Cowling approximation, one can obtain the relation as
\begin{equation}
  \frac{\Delta n_{\rm b}}{n_{\rm b}} = -\left[e^{-\Lambda}W' + \ell(\ell+1)V\right]\frac{1}{r^2}Y_{\ell k},   \label{eq:dn}
\end{equation}
where $\Delta n_{\rm b}$ is the Lagrangian perturbation of the baryon number density and the prime denotes the partial derivative with respect to $r$. Assuming the adiabatic perturbations, the Lagrangian perturbations of the pressure ($\Delta p$) and $\Delta n_{\rm b}$ are related to the adiabatic index $\Gamma_1$ via
\begin{equation}
  \Gamma_1 \equiv \left(\frac{\partial \ln p}{\partial \ln n_{\rm b}}\right)_s = \frac{n_{\rm b}}{p}\frac{\Delta p}{\Delta n_{\rm b}},  
         \label{eq:Gamma1}
\end{equation}
while one can get the additional equation from the energy conservation law (or the first law of thermodynamics), i.e.,
\begin{equation}
 \Delta\varepsilon = (\varepsilon + p)\frac{\Delta n_{\rm b}}{n_{\rm b}},   \label{eq:e_con}
\end{equation}
where $\Delta \varepsilon$ denotes the Lagrangian perturbation of the energy density. Since the Lagrangian perturbation of a property $x$, i.e., $\Delta x$, is associated with the Eulerian perturbation ($\delta x$) in the linear analysis, such as $\Delta x = \delta x + \xi^i\partial_i x$, by combining Eqs. (\ref{eq:Gamma1}) and (\ref{eq:e_con}), one can obtain that 
\begin{equation}
  \delta p = c_s^2 \delta\varepsilon + p\Gamma_1 {\cal A}\xi^r,  \label{eq:dp0}
\end{equation}
where $c_s$ is the sound velocity and ${\cal A}$ is the relativistic Schwarzschild discriminant given by 
\begin{gather}
  c_s^2 \equiv \left(\frac{\partial p}{\partial \varepsilon}\right)_s 
          = \frac{\Delta p}{\Delta \varepsilon} 
          = \frac{p\Gamma_1}{\varepsilon + p}, \\
  {\cal A}(r) \equiv \frac{\varepsilon'}{\varepsilon + p} - \frac{p'}{p\Gamma_1} 
          =  \frac{1}{\varepsilon + p}\left(\varepsilon' - \frac{p'}{c_s^2}\right).
\end{gather}
We remark that ${\cal A}$ is a little different from that introduced in Ref. \cite{Finn1988}, where the factor $e^{-\Lambda}$ is also included in the discriminant. We also remark that $c_s$ is determined from the adopted EOS independently of the stellar structure, while ${\cal A}$ is determined only with the stellar structure. With this discriminant ${\cal A}$, the relativistic Brunt-V\"{a}is\"{a}l\"{a} frequency, $f_{\rm BV}$, is given by
\begin{equation}
  f_{\rm BV} = {\rm sgn}({\cal N}^2)\sqrt{|{\cal N}^2|} / 2\pi,
\end{equation}
where ${\cal N}^2$ is given by \cite{Finn1988}
\begin{equation}
  {\cal N}(r)^2 = -\Phi'e^{2\Phi-2\Lambda}{\cal A}(r).
\end{equation}
We remark that the region with ${\cal A}>0$ (${\cal A}<0$), which corresponds to ${\cal N}^2<0$ (${\cal N}^2>0$), is stable (unstable) with respect to the convection. The radial profiles of ${\cal A}$ and $f_{\rm BV}$ at $T_{\rm pb}=48$, 148, 248, and 348 ms are shown in Fig. \ref{fig:BV}. From this figure, one can see that most regions of PNS seem to be stable with respect to the convection, although this may be an original feature in the PNS model obtained in Ref. \cite{KKT2016} in which the 3D hydrodynamic motion rapidly washes out the negative entropy gradient (see the middle panel in Fig. \ref{fig:back}).
One can also see that the absolute value of $f_{\rm BV}$ becomes larger in the earlier phases after core bounce. Now, $\delta p$ and $\delta \varepsilon$ are generally expressed as $\delta p(t,r)Y_{\ell k}$ and $\delta\varepsilon(t,r)Y_{\ell k}$, respectively. Thus, one obtains the following equations for any $\ell$-th perturbations,
\begin{gather}
  \delta\varepsilon = -\frac{\varepsilon + p}{r^2}\left[e^{-\Lambda}W' + \ell(\ell+1)V\right] - \frac{\varepsilon'}{r^2}e^{-\Lambda}W,   
      \label{eq:de} \\
  \delta p = c_s^2\delta\varepsilon +  \frac{p\Gamma_1 {\cal A}}{r^2}e^{-\Lambda}W,  \label{eq:dp}
\end{gather}
where Eq. (\ref{eq:de}) comes from Eqs. (\ref{eq:dn}) and (\ref{eq:e_con}), while Eq. (\ref{eq:dp}) comes from Eq. (\ref{eq:dp0}).

In addition, the perturbed energy-momentum conservation law, i.e., $\nabla_\nu \delta T^{\mu\nu}=0$, gives us the following equations,
\begin{gather}
  \frac{\varepsilon + p}{r^2}e^{-2\Phi}\ddot{W} + e^{-\Lambda}\delta p' + \Phi' e^{-\Lambda}\left(\delta\varepsilon + \delta p\right) = 0, 
     \label{eq:mur}  \\
  \delta p = (\varepsilon + p)e^{-2\Phi}\ddot{V},  \label{eq:muth}
\end{gather}
which correspond to the $r$- and $\theta$-components of the perturbed energy-momentum conservation law. We remark that the $t$-component of the perturbed energy-momentum conservation law is exactly the same as Eq. (\ref{eq:de}). By combining Eqs. (\ref{eq:de}) -- (\ref{eq:muth}) and assuming that $W(t,r)=e^{i\omega t}W(r)$ and $V(t,r)=e^{i\omega t}V(r)$, one can get the perturbation equations for $W$ and $V$ as
\begin{gather}
  W' = \frac{1}{c_s^2}\left(\Phi'W + \omega^2 r^2e^{-2\Phi+\Lambda}V\right) - \ell(\ell+1)e^{\Lambda}V, \\
  V' = -\frac{1}{r^2}e^{\Lambda}W + 2\Phi' V - {\cal A}\left(\frac{1}{\omega^2r^2}\Phi'e^{2\Phi-\Lambda}W + V\right).
\end{gather}
In order to solve this equation system, one has to impose appropriate boundary conditions. The regularity condition should be imposed at the stellar center, i.e.,
\begin{gather}
  W = W_0 r^{\ell+1} \ \ {\rm and}\ \  V=-\frac{W_0}{\ell}r^{\ell},
\end{gather}
where $W_0$ is constant. The boundary condition is that the Lagrangian perturbation of pressure should be zero at the surface of PNS, i.e., 
\begin{equation}
  \Phi'e^{-\Lambda}W + \omega^2 r^2 e^{-2\Phi}V = 0,
\end{equation}
for the case that the PNS surface is determined by the critical density, while it is that the radial displacement should be zero at the shock radius, i.e., $W=0$, for the case that the oscillations are considered in the domain inside the shock radius.  At last, the problem to solve becomes the eigenvalue problem with respect to $\omega$. Once the eigenfrequency $\omega$ is determined, it is connected to the oscillation frequency, $f$, via $f=\omega/2\pi$.

\section{Asteroseismology of PNS}
\label{sec:IV}

\begin{figure}[htbp]
\begin{center}
\includegraphics[scale=0.5]{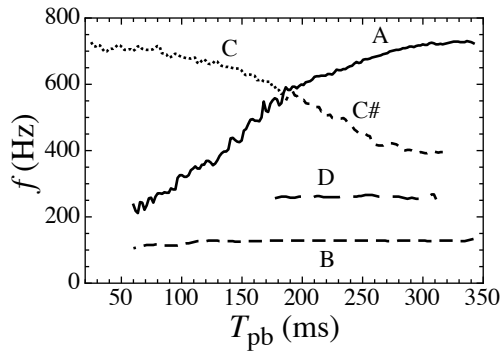}
\end{center}
\caption{
The characteristic GW frequencies extracted by the  time-frequency analysis \cite{Kawahara2018} for the 3D model SFHx in \cite{KKT2016}.
}
\label{fig:Kawahara}
\end{figure}

Recently, the sophisticated time-frequency analysis \cite{Kawahara2018} showed that the various GW signatures with wide frequency ranges can be extracted from the GW spectrogram for the 3D-GR model (SFHx) employed in this work \cite{KKT2016}. For instance, as shown in Fig. \ref{fig:Kawahara}, they found the sequences of A, B, C, C\#, and D. The sequence A has been observed in the several previous studies, which is considered as ``the surface $g$-mode" \cite{MJM2013,CDAF2013,KKT2016} of the PNS. The sequences B and D could come from the mass accretion influenced by SASI \cite{KKT2016}. It is noteworthy that the low frequency component B has been also reported in other recent 3D studies \cite{OC2018,Andresen18}. The excitation mechanism of the sequence C (and C\#) is still unclear. In this paper, we attempt to compare the eigenfrequencies derived from the perturbation analysis with
 the GW frequencies obtained from the hydrodynamics simulations as in Fig. \ref{fig:Kawahara}. As a baseline, we mainly focus on the sequence A in this work.

\begin{figure}[htbp]
\begin{center}
\includegraphics[scale=0.5]{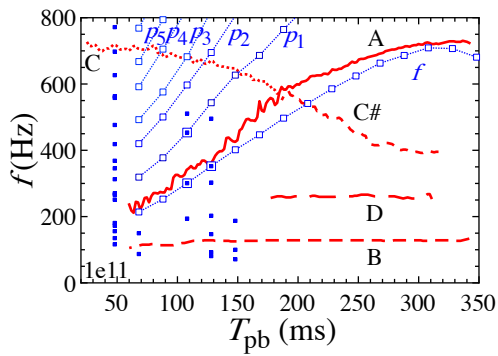} 
\end{center}
\caption{
Eigenfrequencies in the PNS model with $\rho_s=10^{11}$ g cm$^{-3}$. In particular, the $f$ and $p_i$ for $i=1-6$ are explicitly shown with the open-squares together with the dotted lines, where the double squares denote the cases that the node number in the eigenfunction is different from the standard definition (see Fig. \ref{fig:wf-1e11-128}). For reference, the various excited GW frequencies derived from the simulation data are also shown with the red lines. 
}
\label{fig:mode-1e11}
\end{figure}

\begin{figure}[htbp]
\begin{center}
\begin{tabular}{cc}
\includegraphics[scale=0.46]{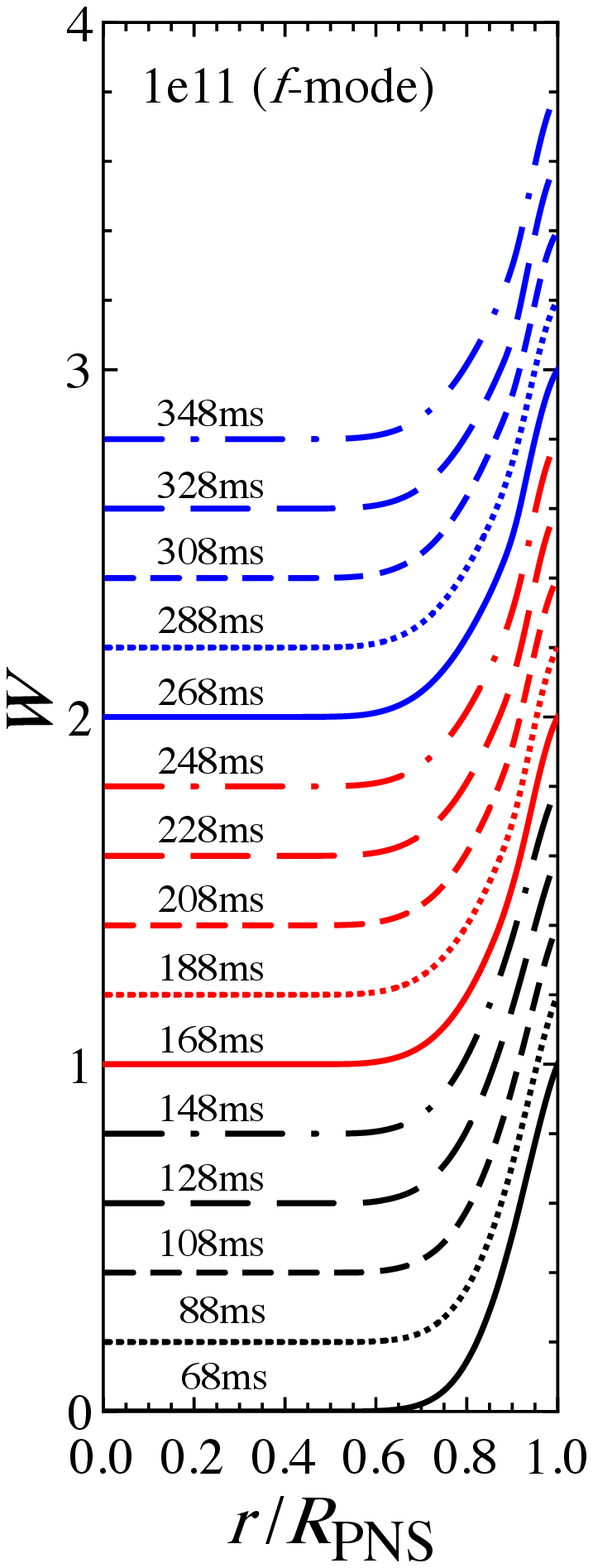} &
\includegraphics[scale=0.46]{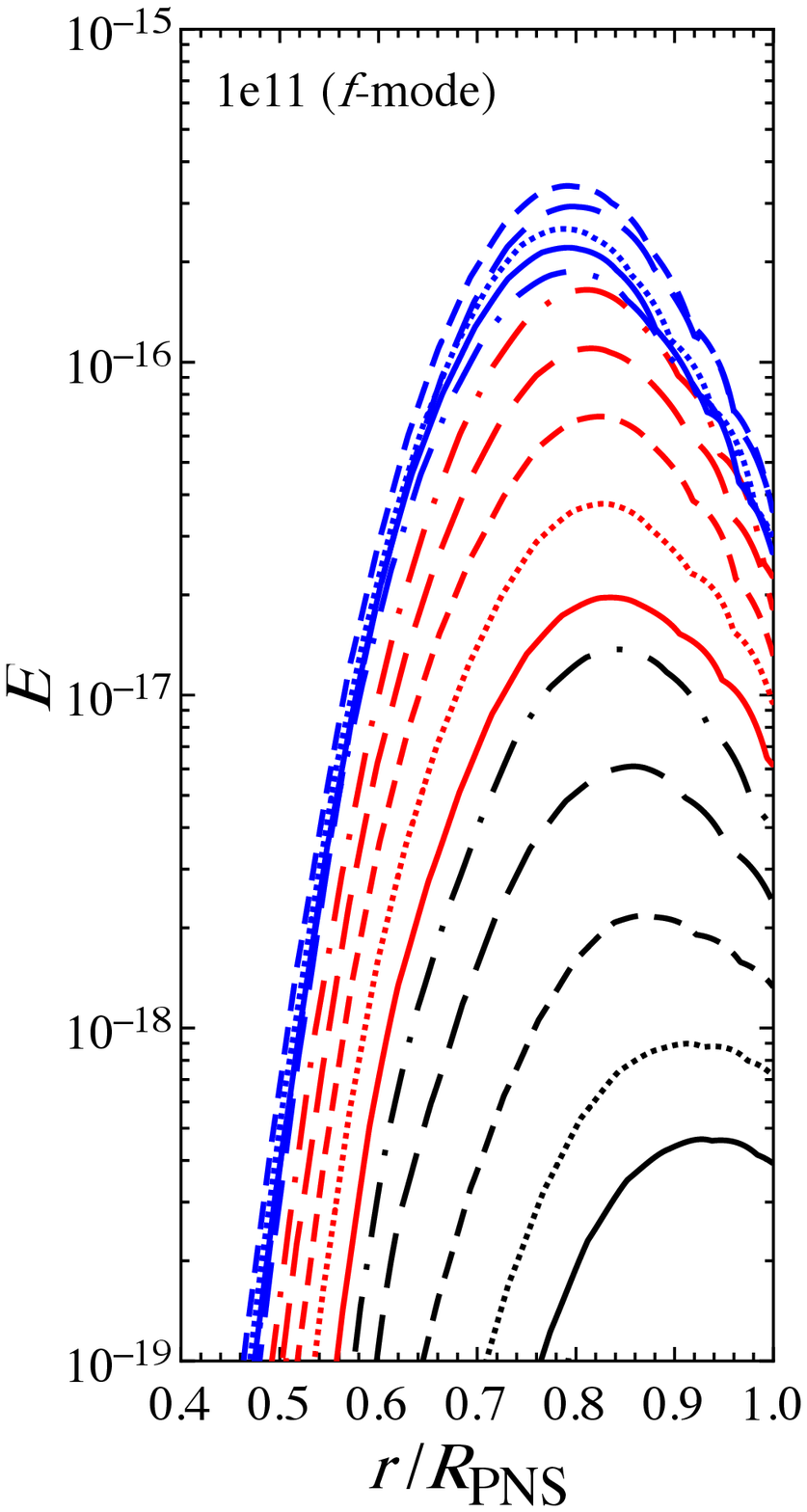} 
\end{tabular}
\end{center}
\caption{
The eigenfunctions of $W$ for the $f$-modes are shown in the left panel, where the amplitude of $W$ is normalized by the amplitude at the stellar surface and is shifted a little for easily distinguishing the lines. The radial-dependent pulsation energy  density $E$ given by Eq. (\ref{eq:energy}) is shown in the right panel, where the amplitudes are normalized appropriately.
}
\label{fig:wE-1e11}
\end{figure}

\subsection{PNS surface determined by the fixed density}
\label{sec:IV-A}

\begin{figure}[htbp]
\begin{center}
\includegraphics[scale=0.5]{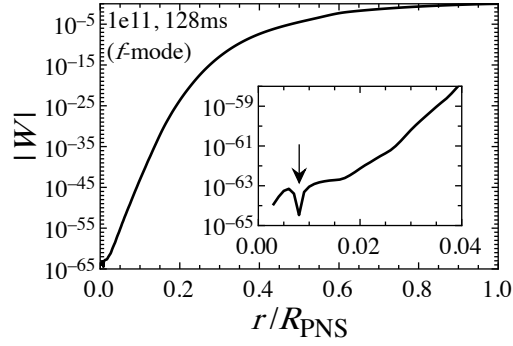} 
\end{center}
\caption{
The details of the eigenfunction of $|W|$ for the $f$-modes at 128ms for the PNS model with $\rho_s=10^{11}$ g/cm$^3$ is shown, which is an example of the specific case shown with the double squares in Fig. \ref{fig:mode-1e11}. This sample has one node in the eigenfunction shown by the arrow, but the shape of eigenfunction is almost the same as that of the other $f$-mode shown in the left panel of Fig. \ref{fig:wE-1e11}. 
}
\label{fig:wf-1e11-128}
\end{figure}

\begin{figure}[htbp]
\begin{center}
\includegraphics[scale=0.5]{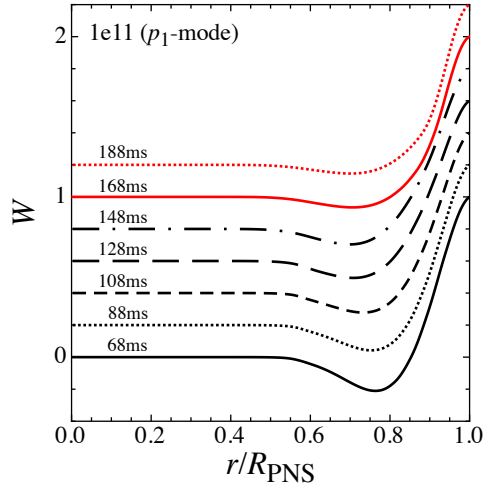} 
\end{center}
\caption{
The eigenfunctions of $W$ for the $p_1$-modes are shown for the PNS model with $\rho_s=10^{11}$ g/cm$^3$, where the amplitude is normalized by the surface amplitude and is shifted a little.
}
\label{fig:wp1-1e11}
\end{figure}

First, we consider the PNS model, whose surface density ($\rho_s$) is changed in a parametric manner. In Fig. \ref{fig:mode-1e11}, we show the eigenfrequencies determined in the PNS model with $\rho_s=10^{11}$ g cm$^{-3}$. Among many eigenfrequencies, we identify the $f$ and $p_i$-modes for $i=1-6$, which are shown with open squares (and double squares) connected by thin-dotted lines. In addition, for reference the various excited GW frequencies derived from the simulation data are shown in Fig. \ref{fig:Kawahara}. The eigenfunctions of $W$ for the $f$-modes with various time steps are shown in the left panel of Fig. \ref{fig:wE-1e11}, where the amplitude of $W$ is normalized by the value at the PNS surface and is shifted upward in order to distinguish each line easily. From this figure, one can observe that the eigenfunctions of $W$ at any time step become as the standard definition of $f$-mode, i.e., the eigenfunction monotonically increases outward without any nodes. However, in fact, the eigenfunctions at 108 and 128 ms are different from the standard definition, where the node number is more than one. Even so, since the both eigenfunctions are very similar to the other $f$-mode eigenfunctions, we identify these modes as $f$-modes. For example, the eigenfunction of $|W|$ at 128 ms is shown in Fig. \ref{fig:wf-1e11-128}. In the similar way, by checking the shape of the eigenfunctions and by counting the node number, we identify the $p_1$- and $p_2$-modes as shown in Fig. \ref{fig:mode-1e11},  where the eigenfunctions of $W$ for the $p_1$-modes at each time step are shown in Fig. \ref{fig:wp1-1e11}. Again, the double squares denote the eigenfrequencies that are different from the standard definition by the node number.

\begin{figure}[htbp]
\begin{center}
\includegraphics[scale=0.5]{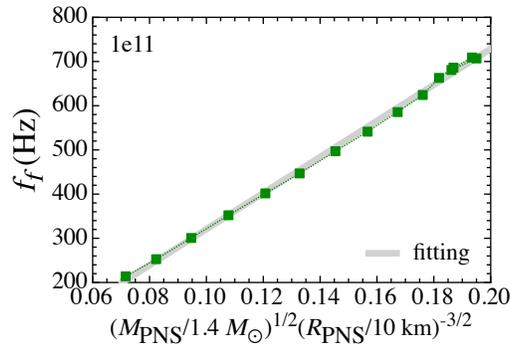} 
\end{center}
\caption{
For the PNS model with $\rho_s=10^{11}$ g/m$^3$, the $f$-mode frequencies are shown as a function of the square root of the PNS average density. The solid line denotes the fitting formula given by Eq. (\ref{eq:fit}), while the dashed line is the analytical formula of the $f$-mode frequency for the star with uniform incompressible fluid given by Eq. (\ref{eq:ana-f}). 
}
\label{fig:fave-1e11}
\end{figure}

In the right panel of Fig. \ref{fig:wE-1e11}, we also show the radial dependent pulsation energy density in the $f$-mode oscillation at each time step, where in the same way as in Refs. \cite{TCPF2018,MRBV2018}, the Newtonian pulsation energy density at each radial position can be estimated with our variables as 
\begin{equation}
  E(r) \sim \frac{\omega^2\varepsilon}{r^4} \left[W^2 + \ell(\ell+1)r^2V^2\right].   \label{eq:energy}
\end{equation}
One can observe that the amplitude of the $f$-mode eigenfunction becomes maximum at the stellar surface. The pulsation energy density, however, becomes maximum at around $80-90 \%$ of the PNS radius.  From Fig. \ref{fig:mode-1e11} we can see a good agreement of the sequence A (which is referred as the surface $g$-mode \cite{MJM2013,CDAF2013,KKT2016}) with the $f$-mode oscillations in the PNS, when we take the specific surface density of $10^{11}$ g cm$^{-3}$.  In Fig. \ref{fig:mode-1e11}, the filled squares correspond to the eigenfrequencies, which we can not unambiguously identify as either $f$-, $p_1$-, or $p_2$-modes (e.g., Fig.~\ref{fig:w-1e10-108}).
   These modes are left unidentified in this work. 

\begin{figure*}[htbp]
\begin{center}
\begin{tabular}{cc}
\includegraphics[scale=0.5]{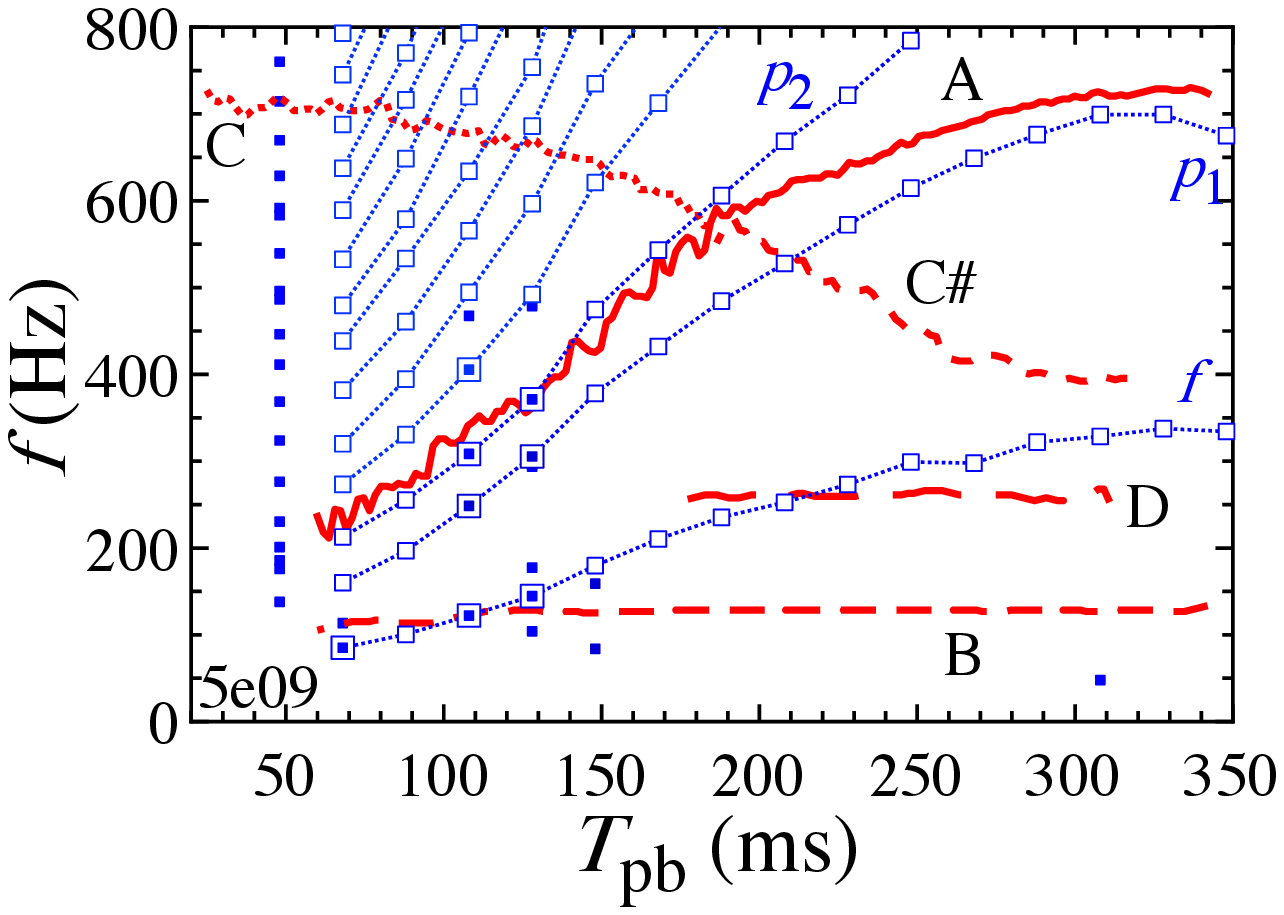} &
\includegraphics[scale=0.5]{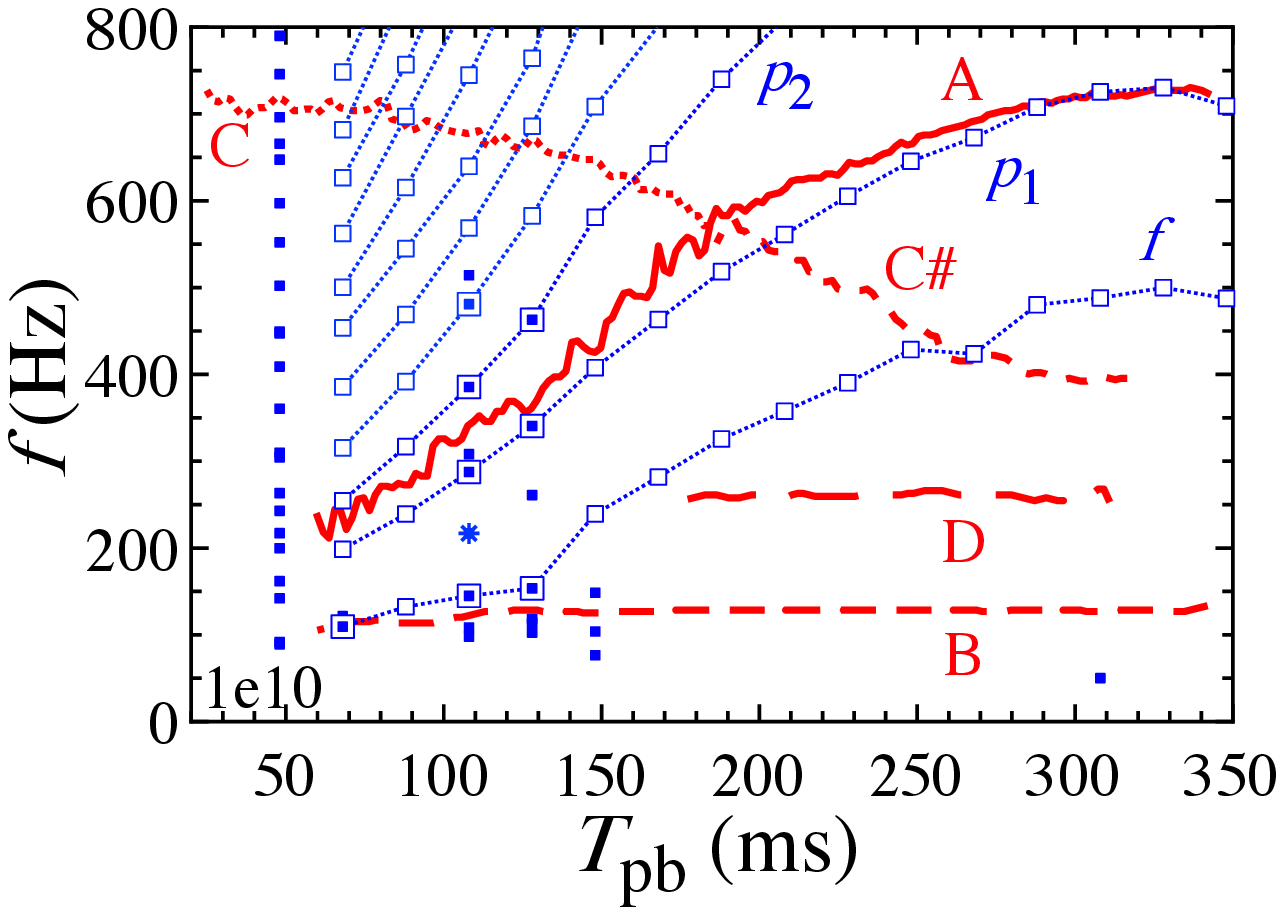} 
\end{tabular}
\end{center}
\caption{
Same as in Fig. \ref{fig:mode-1e11}, but for the PNS models with $\rho_s=5\times 10^9$ (left panel) and $10^{10}$ g cm$^{-3}$ (right panel).   For the PNS model with $\rho_s=10^{10} $ g/cm$^3$, as an example of the eigenfunction of which eigenmode is left 
unidentified, we will consider the frequency at 108 ms shown by the asterisk, discussed with Fig. \ref{fig:w-1e10-108}.
}
\label{fig:modes}
\end{figure*}

\begin{figure}[htbp]
\begin{center}
\includegraphics[scale=0.6]{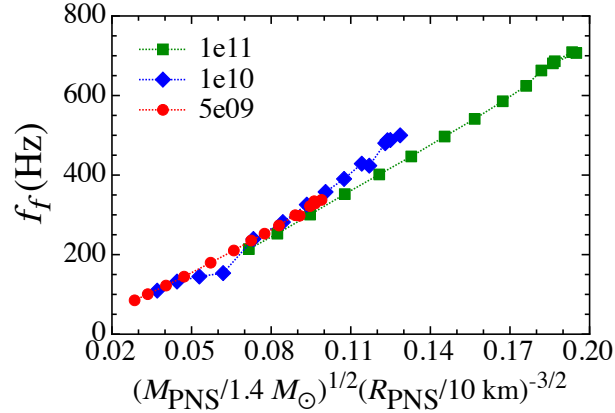} 
\end{center}
\caption{
The $f$-mode frequencies from the PNS models with different definition of the surface density are shown as a function of the corresponding PNS average density, where the squares, diamonds, and circles correspond to the results with $\rho_s=10^{11}$, $10^{10}$, and $5\times 10^{9}$ g cm$^{-3}$, respectively. 
}
\label{fig:fave}
\end{figure}

\begin{figure}[htbp]
\begin{center}
\begin{tabular}{cc}
\includegraphics[scale=0.46]{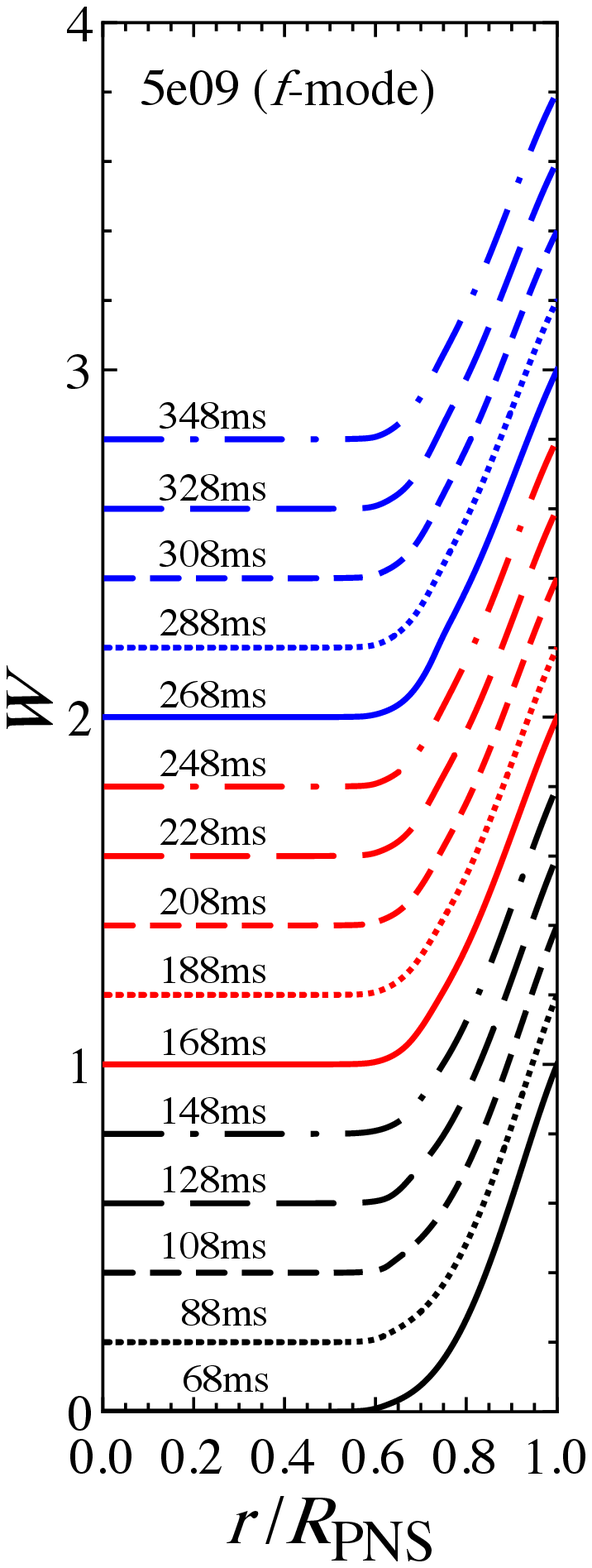} &
\includegraphics[scale=0.46]{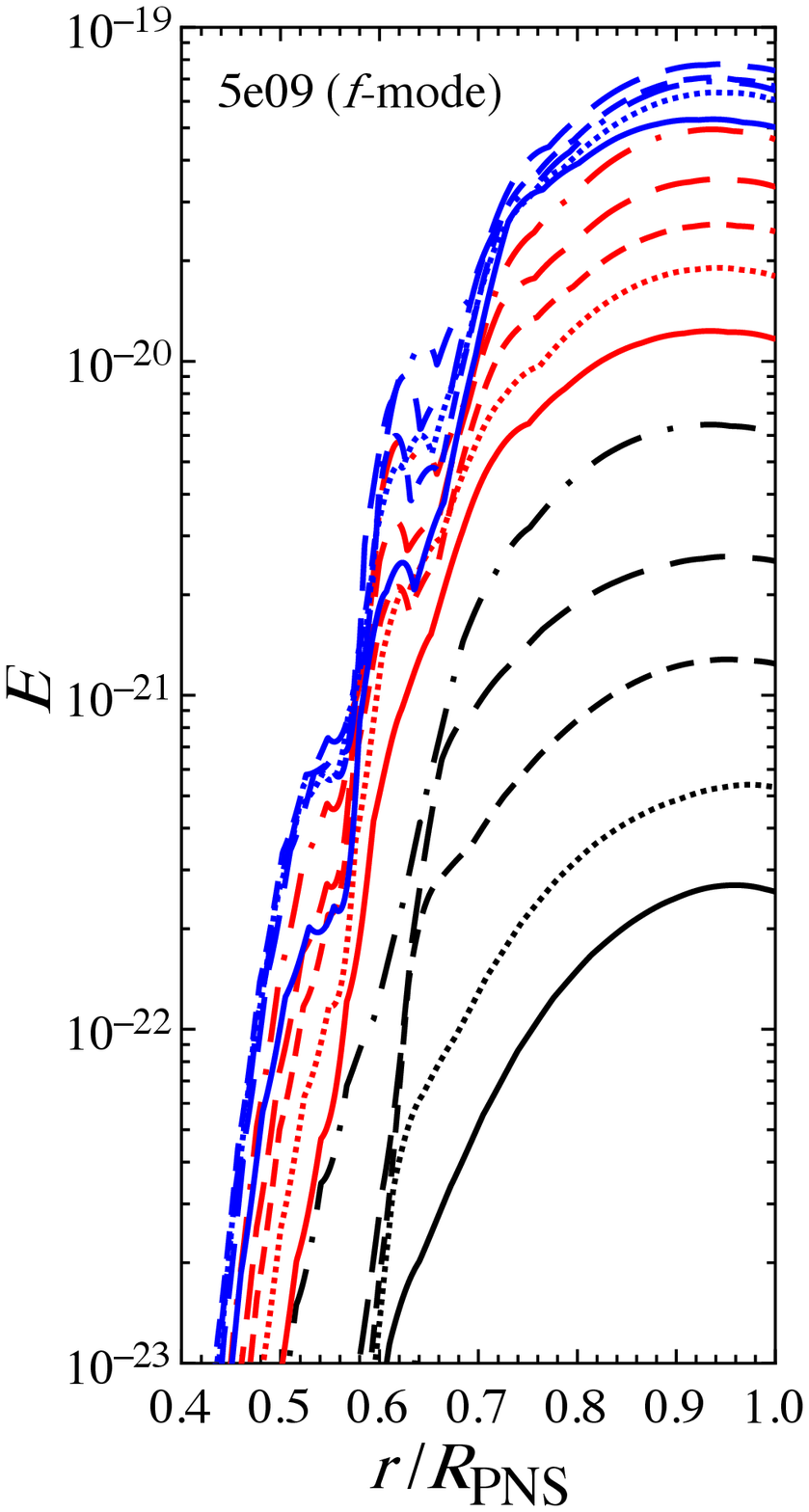} 
\end{tabular}
\end{center}
\caption{
Same as in the left panel of Fig. \ref{fig:wE-1e11}, but for the PNS models with $\rho_s=5\times 10^9$ g/cm$^3$. 
}
\label{fig:wf}
\end{figure}

\begin{figure}[htbp]
\begin{center}
\includegraphics[scale=0.6]{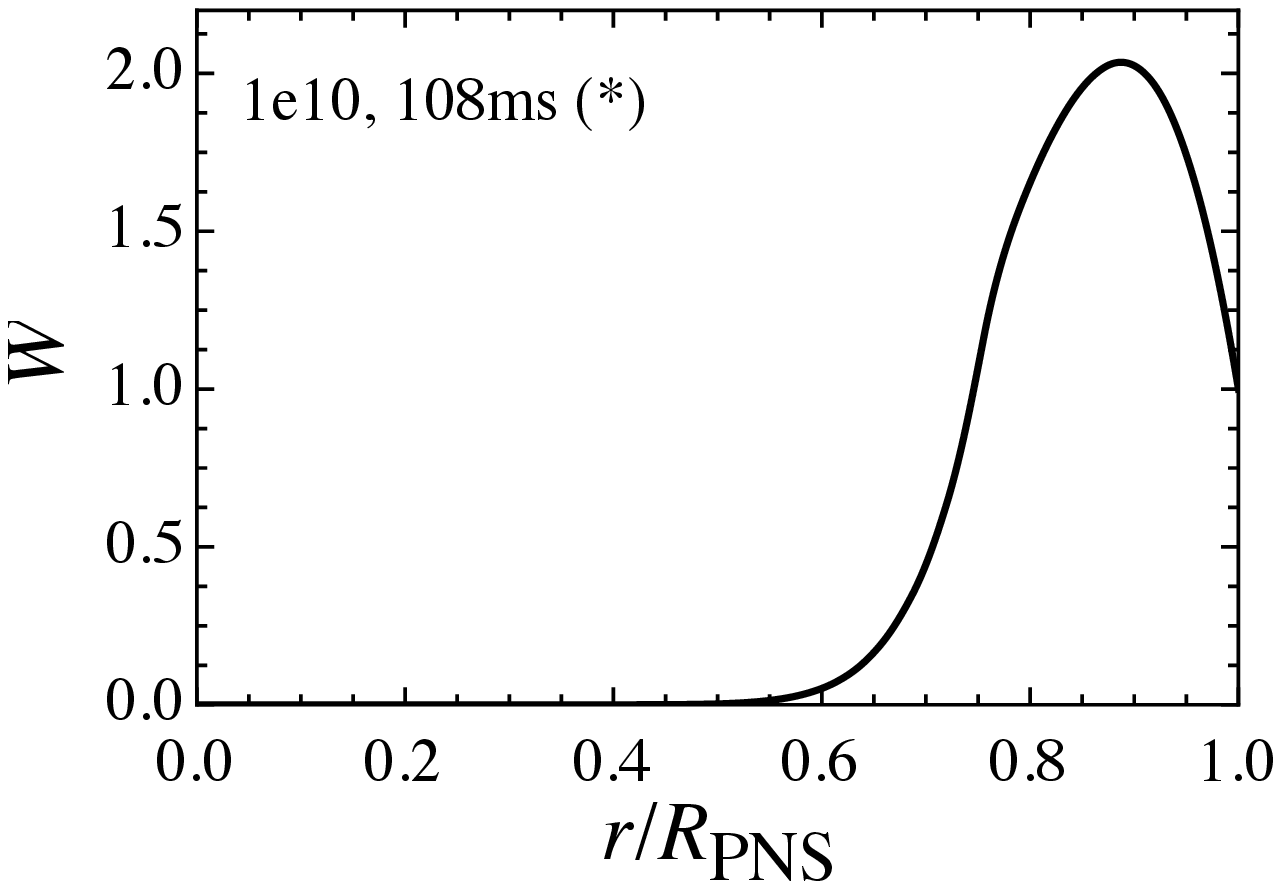} 
\end{center}
\caption{
An example of the eigenfunction $W$, of which eigenmode is left unidentified (from the radial node number of $W$).  
This corresponds to the frequency shown by the asterisk in the right panel of Fig.~\ref{fig:modes}, which is close to the sequence A.
}
\label{fig:w-1e10-108}
\end{figure}

The identification of the sequence A with the $f$-mode oscillation
 indicates that one could extract the PNS properties by observing the $f$-mode originated GWs. In practice, it is well-known that, since the $f$-mode is associated with the acoustic waves, its frequency can be characterized with the stellar average density almost independently of the adopted EOS (see \cite{AK1996,AK1998} for cold NSs and also \cite{ST2016,SKTK2017} even for the PNS models). In Fig. \ref{fig:fave-1e11}, we show the $f$-mode frequency from the PNS model for each time step as a function of the corresponding square root of the PNS average density. From this figure, as in the previous studies, one can observe that the $f$-mode frequencies can be expressed as a linear function of the square root of the PNS average density. Additionally, with this data, we obtain the fitting formula expressing the $f$-mode frequency, i.e.,
\begin{equation}
  f_f\ {\rm (Hz)} = -87.34 + 4080.78 \left(\frac{M_{\rm PNS}}{1.4M_\odot}\right)^{1/2}\left(\frac{R_{\rm PNS}}{10\ {\rm km}}\right)^{-3/2},
    \label{eq:fit}
\end{equation}
 where $M_{\rm PNS}$ and $R_{\rm PNS}$ denotes the PNS (gravitational) mass and radius, respectively.
The resultant fitting formula is also shown in Fig. \ref{fig:fave-1e11} with the thick-solid line. With this fitting formula, one may know the time evolution of the PNS average density from the observation of the gravitational waves. We remark that the $\ell$-th $f$-mode frequency for the star with uniform incompressible fluid has been derived analytically as
\begin{equation}
 f_f^{(\rm a)} = \frac{1}{2\pi}\sqrt{\frac{2M_{\rm PNS}}{R_{\rm PNS}^3}\frac{2\ell(\ell-1)}{2\ell+1}}, \label{eq:ana-f}
\end{equation}
which is known as a Kelvin $f$-mode. The expected $\ell=2$ frequency is also shown in Fig.~\ref{fig:fave-1e11} with dashed line, but it seems that this formula assuming the incompressible fluid is not suitable for expressing the $f$-mode frequencies for the PNS models.

In the similar way, we determine the eigenfunctions in the PNS models with $\rho_s=5 \times 10^9$ and $10^{10}$ g/cm$^3$, which are shown in Fig. \ref{fig:modes}. From this figure together with Fig. \ref{fig:mode-1e11}, we find that the eigenfrequencies depend on the selection of the surface density of the PNS model. In fact, the frequencies of $f$ and $p_i$-modes decrease, as $\rho_s$ decreases. This tendency may be understood as a result of the decrease of the average density of the PNS, as $\rho_s$ decreases, because the $f$ and $p_i$-modes are a kind of acoustic waves, whose frequencies can be characterized by the average density of the PNS. In fact, as shown in Fig. \ref{fig:fave}, the $f$-mode frequencies can be expressed well as a function of the PNS average density, even though the definition of the surface density is different.

 The dependence of the eigenfrequencies on the surface density seems to be consistent with Ref. \cite{MRBV2018} as least in the early postbounce phase. On the other hand, in the phase later than $\sim 500$ ms after bounce, Morozova et al. \cite{MRBV2018} showed that the eigenfrequencies are almost independent from the selection of the surface densities of the PNS models. This could be because the density gradient in the vicinity of the background PNS models becomes steeper in the later phase, making the average density less sensitive to the choice of the surface densities. 
Thus, although the GW signal (the sequence A) obtained in the 3D numerical simulation \cite{KKT2016} is well ascribed to the $f$-mode oscillations in the PNS model with $\rho_s=10^{11}$ g/cm$^3$, this result may not be universal at least in the early postbounce phase, i.e., one may have to select a specific surface density to identify the GW signal. In such a case, it could be more difficult to extract the PNS information from direct GW observations.

Additionally, for the PNS model with $\rho_s=5 \times 10^9$ g/cm$^3$, the eigenfunction of $W$ and the radial dependent pulsation energy density at each time step are shown in Fig.~\ref{fig:wf}.  We remark that eigenfunction of $W$ and the radial dependent pulsation energy density for the PNS model with $\rho_s=10^{10}$ g/cm$^3$ are more or less similar to those for the PNS model with $\rho_s=5 \times 10^9$ g/cm$^3$. The eigenfunctions of $W$ look similar to those shown in Fig. \ref{fig:mode-1e11}, but one can see the difference in the radial dependent pulsation energy density. From this figure, it seems that the oscillations around the stellar surface become more important in the PNS model with lower $\rho_s$. 
 Furthermore, as an example of the eigenmode that could not be identified as a specific mode, we show the eigenfunction of $W$ for the PNS model with $\rho_s=10^{10}$ g/cm$^3$ at 108 ms, which is shown with the asterisk in the right panel of Fig. \ref{fig:modes}. Obviously, this eigenfunction is satisfied the boundary condition but the shape of eigenfunction is apparently different from the other $f$- or $p_1$ mode.  We  note that the lower frequencies, such as sequences B or D in Fig. \ref{fig:Kawahara}, are not excited in the PNS models with the specific surface density irrespective of its value. In addition, some of the eigenfrequencies lower than the $f$-mode in Figs. \ref{fig:mode-1e11} and \ref{fig:modes} could be considered as $g$-mode oscillations. However these modes are left unidentified because of the lack of the clear node structure in the eigenfunctions as mentioned above.

\begin{figure}[tbp]
\begin{center}
\begin{tabular}{cc}
\includegraphics[scale=0.5]{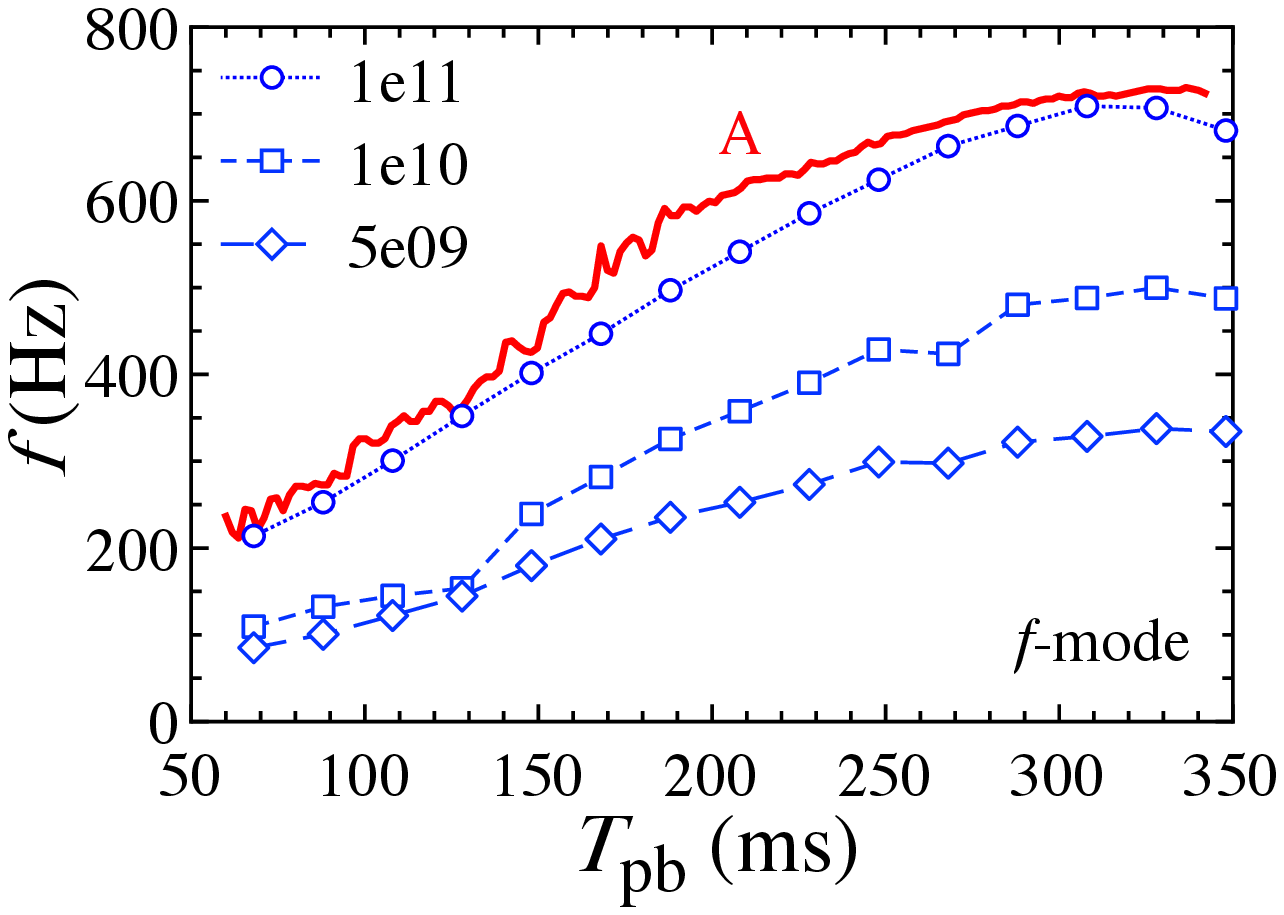} & 
\includegraphics[scale=0.5]{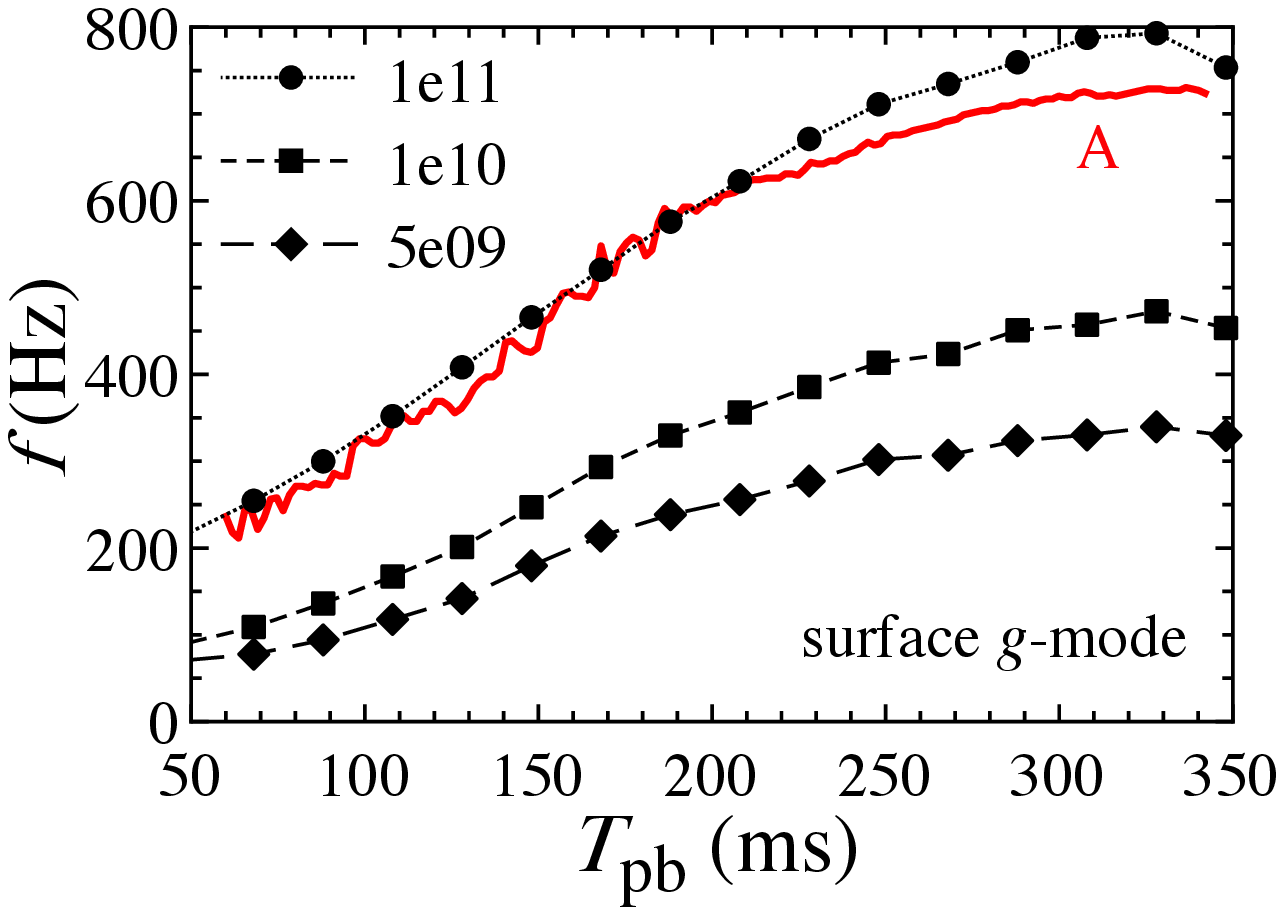}
\end{tabular}
\end{center}
\caption{The GW signal (the sequence of A) is compared with the $f$-mode GW of PNS in the left panel and with the surface $g$-mode calculated with Eq. (\ref{eq:fpeak}) in the right panel, where the circles, squares, and diamonds denote the PNS models constructed with $\rho_s=10^{11}$, $10^{10}$, and $5\times 10^{9}$ g/cm$^3$, respectively.
}
\label{fig:comparison}
\end{figure}

 Finally, the GW signal (the sequence A) is compared with the $f$-mode frequencies calculated in this study with different surface density and the surface $g$-mode with the formula proposed in Ref.~\cite{MJM2013}, i.e.,
\begin{equation}
  f_{\rm peak} = \frac{1}{2\pi}\frac{M_{\rm PNS}}{R_{\rm PNS}^2}\sqrt{\frac{1.1 m_n}{\langle E_{{\bar{\nu}}_e}\rangle}}\left(1-\frac{M_{\rm PNS}}{R_{\rm PNS}}\right)^2,  \label{eq:fpeak}
\end{equation}
where $\langle E_{{\bar{\nu}}_e}\rangle$ denotes the mean energy of electron antineutrinos and $m_n$ is the neutron mass. We remark that the Brunt-V\"{a}is\"{a}l\"{a} frequencies estimated at the PNS surface is original ``surface $g$-mode", with which Eq. (\ref{eq:fpeak}) is approximately derived. Those frequencies are shown in Fig. \ref{fig:comparison}, where the left and right panels correspond to the results of $f$-mode frequencies obtained in the linear analysis and surface $g$-mode frequencies calculated with Eq. (\ref{eq:fpeak}), respectively, for the PNS models with $\rho_s=10^{11}$ (circles), $10^{10}$ (squares), and $5\times 10^{9}$ g/cm$^3$ (diamonds). From this figure, one can observe that the both frequencies strongly depend on the surface density, but agree well with the GW signal of the sequence of A for the PNS model with $\rho_s=10^{11}$ g/cm$^3$. Even so, since the surface $g$-mode (or the Brunt-V\"{a}is\"{a}l\"{a} frequency at the PNS surface) is the local value while $f$-mode is the global oscillations of PNS, it may be more natural that the GW signal (sequence of A) is considered as a result of the $f$-mode oscillations.

\subsection{PNS inside the shock radius}
\label{sec:IV-B}

Next, we consider the oscillations inside the shock radius. In this case, as mentioned before, the boundary condition at the shock radius is that the radial component of the Lagrangian displacement should be zero. That is, the eigenfunction of $W$ is always zero at the shock radius, where the standard classification of the eigenmode may not be adopted. Intrinsically, the eigenvalue problem to solve with this PNS model is significantly different from that with the PNS model whose surface density is fixed. Anyway, as an advantage of this PNS model, the ambiguity for selecting the position of boundary disappears, while the spherical symmetric model may not be a good assumption in the region whose density is very low, because the matter motion is not neglected in such a region.  Furthermore the excitation of GWs in the numerical simulation may come from such an oscillation inside the whole shocked region, although there are currently a few studies \cite{TCPF2018,TCPOF2019} examining this effect.

\begin{figure}[tbp]
\begin{center}
\includegraphics[scale=0.5]{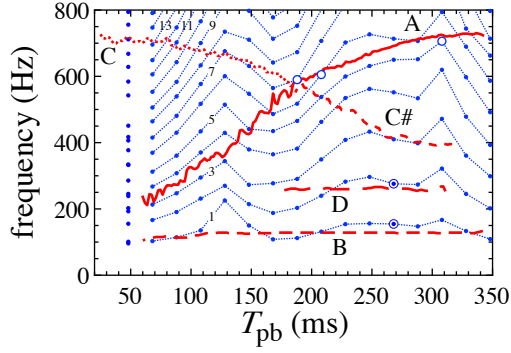}
\end{center}
\caption{
The eigenfrequencies calculated in this study are shown with marks, while the excited GW frequencies in the numerical simulation are again shown with various red lines. The double circles are the lowest and the second lowest eigenfrequencies at 268 ms, which are focused in Fig. \ref{fig:268ms}. The open circles are the examples, of which eigenfunctions will be discussed in Fig. \ref{fig:wshock}. The modes, whose eigenfunctions are the similar to each other, are connected with the dotted lines.  
}
\label{fig:shock}
\end{figure}

We try to determine the eigenfrequencies with the PNS model inside the shock radius. The resultant eigenfrequencies are shown in Fig. \ref{fig:shock}, where the same modes are connected with dotted lines. From this figure, one can observe that even lower frequencies can be excited with the PNS model inside the shock radius, which is different feature compared to the results with the PNS model whose surface density is fixed. In fact, in some time interval, it seems that the eigenfrequencies are excited close to the sequences B and D.

\begin{figure*}[tbp]
\begin{center}
\begin{tabular}{cc}
\includegraphics[scale=0.5]{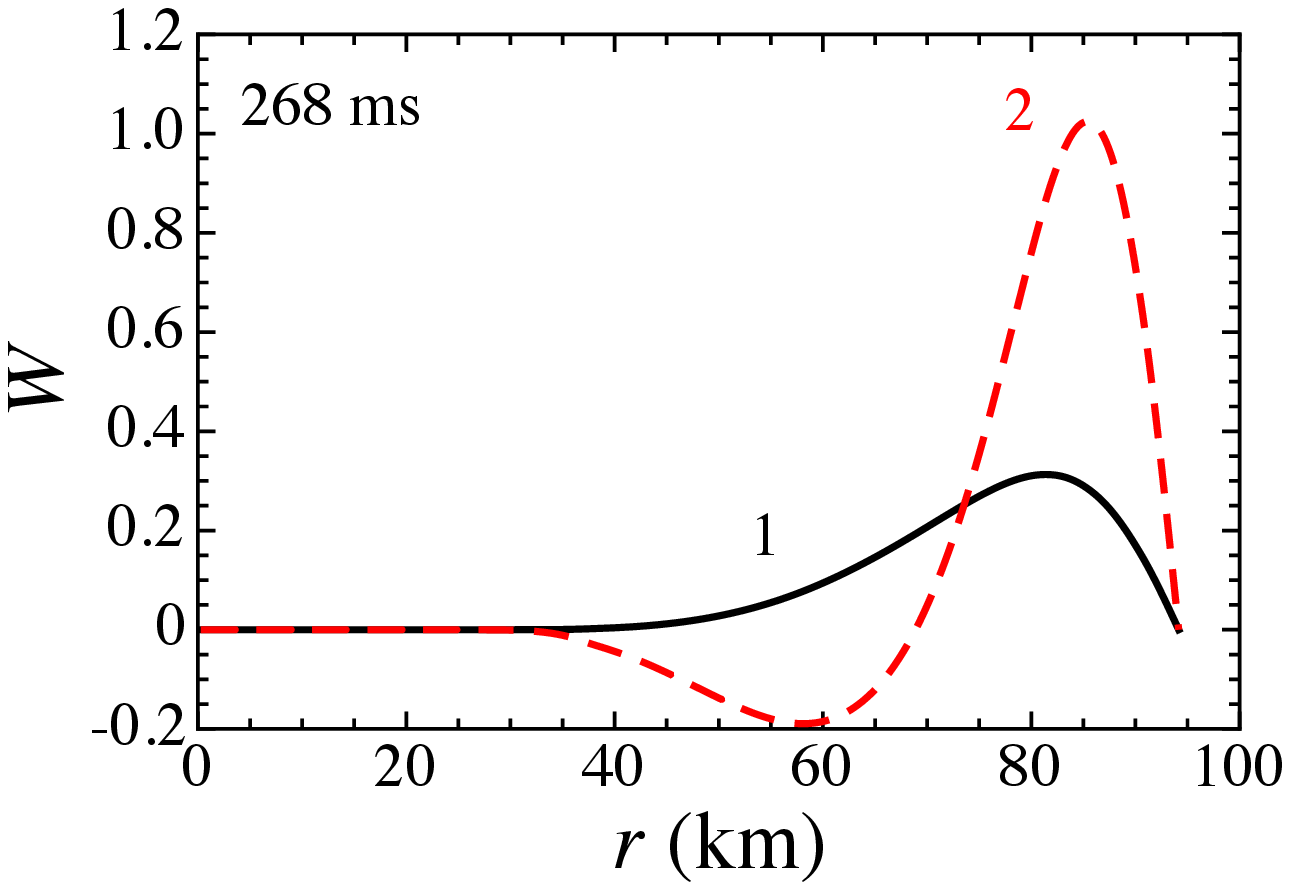} &
\includegraphics[scale=0.5]{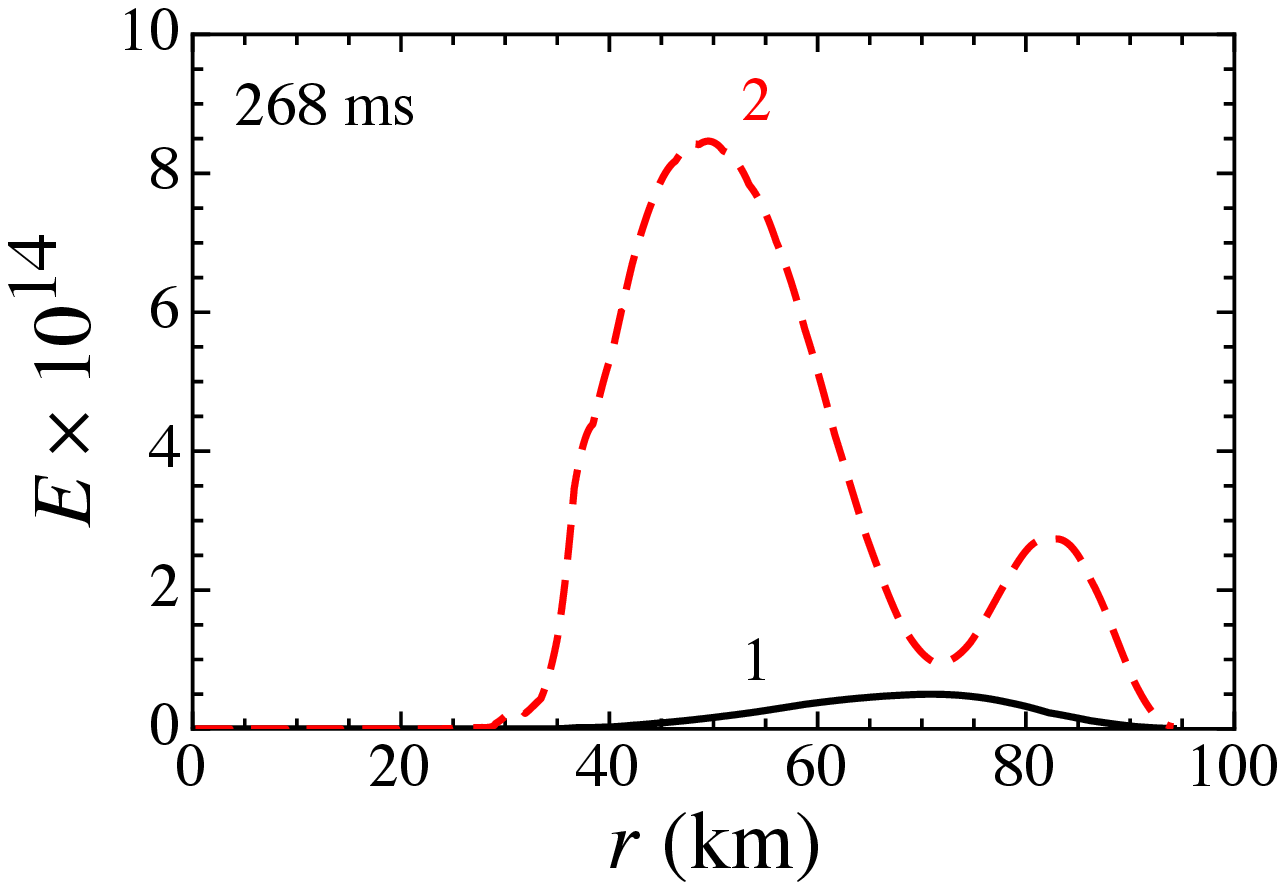}  
\end{tabular}
\end{center}
\caption{
The left and right panels correspond to the eigenfunction $W$ and pulsation energy density given by Eq. (\ref{eq:energy}) for the lowest (denoted with 1) and the second lowest (denoted with 2) eigenfrequencies at 268 ms after core bounce, respectively. In the both panels, the vertical axis is normalized appropriately.
}
\label{fig:268ms}
\end{figure*}

In order to see the oscillation behavior for such eigenfrequencies, we especially focus on the lowest and the second lowest eigenfrequencies at 268 ms after core bounce, which are denoted with the double circles in Fig. \ref{fig:shock}. The corresponding eigenfunction of $W$ and the radial dependent pulsation energy density are shown in Fig. \ref{fig:268ms}, where the solid and dashed lines correspond to the results with the lowest and the second lowest eigenfrequencies, respectively. One can see that the eigenfunctions are very similar to the standard classification of stellar oscillation except for the behavior close to the stellar surface, i.e., the lowest and the second lowest eigenfrequencies may correspond to the $f$- and $p_1$-modes. In addition, in the similar fashion to the results with the PNS model whose surface density is fixed, the amplitude of eigenfunction $W$ and the radial dependent pulsation energy density become significant on the outer part of the oscillation region. However, it has been reported that the excitation of the GW signal according to the sequence B (or maybe also D) effectively comes from the inner part of the PNS, such as $\sim 20$ km \cite{KKT2016,Andresen16}. Thus, although the lower eigenfrequencies obtained via the eigenvalue problem inside the shock radius appear close to the sequences B and D, these frequencies may not physically correspond to the excitation of gravitational wave signal in the sequences B and D. 
We remark that in our model we found only $f$- and $p_i$-mode like frequencies, while not only $f$- and $p_i$-mode like frequencies but also $g_i$-mode like frequencies are found in the previous similar analysis \cite{TCPF2018,TCPOF2019}. This discrepancy may come from the different PNS models obtained by the numerical simulations. We need further studies changing the PNS models to draw a robust conclusion on this choice of the boundary condition.

\begin{figure}[tbp]
\begin{center}
\includegraphics[scale=0.5]{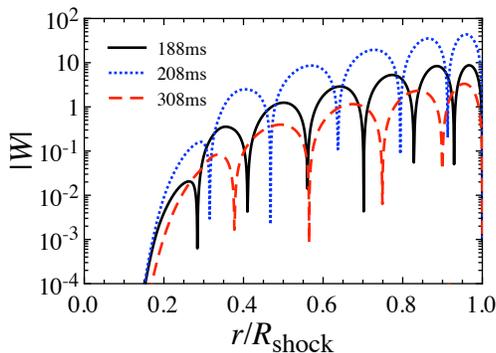}
\end{center}
\caption{
Examples of the eigenfunctions of $|W|$ with the frequencies shown by the open circles in Fig. \ref{fig:shock}.
}
\label{fig:wshock}
\end{figure}

There are a few possible reasons to explain the discrepancy between the the sequence B (and D) and the eigenmodes obtained by the linear analysis. The first and perhaps the main reason is that, in our perturbation analysis using the static background model, the restoring force against the perturbations is assumed to be the acoustic mode. On the contrary, the SASI that is considered to be the emission mechanism of the component B \cite{KKT2016,Andresen16} is sustained by the cycle of the fluid advection and the acoustic mode \cite{Foglizzo06}. It may thus not be suitable to use the static background model that completely omits the fluid advection. As the second reason, the background model is actually far from spherical symmetry particularly in the non-linear SASI phase ($T_{\rm pb}\gtrsim150$ ms, \cite{KKT2016}). These facts would make it even harder to extract the proper eigenmodes for the corresponding GW components.
We also remark that one can not clearly find a specific correspondence between the eigenfrequencies and the sequence A on Fig.~\ref{fig:shock}. 
 In practice, we show examples of the eigenfunction of $|W|$ in Fig.~\ref{fig:wshock} with the frequencies shown by the open circles in Fig.~\ref{fig:shock}, which are close to the sequence A, but one can not straightforwardly identify these modes as the same eigenmode by checking the shape of the eigenfunction (or the radial node numbers). Anyway, in this study we have made a linear analysis with only one result obtained by the numerical simulation in Ref. \cite{KKT2016}, i.e., our result may not be always acceptable for any PNS models. In order to make a robust statement for the CCSN GW signals, we have to make more systematical analyses somewhere by adopting various results of different numerical simulations.

\section{Conclusion and Discussions}
\label{sec:V}
In an attempt to obtain the eigenfrequencies of a PNS in the postbounce phase of CCSNe, we have performed a linear perturbation analysis of the angle-averaged PNS profiles using results from a general relativistic 
CCSN simulation of a $15 M_{\odot}$ star. Particularly, we paid attention to how the choice of the outer boundary condition could affect the PNS oscillation modes in the linear analysis.
By changing the density at the outer boundary of the PNS surface in a parametric manner, we showed that the eigenfrequencies strongly depend on the surface density. By comparing with the GW signals from the hydrodynamics model,  it was shown that the so-called surface $g$-mode of the PNS can be well ascribed to the fundamental oscillations of the PNS. The best match was obtained when the 
PNS surface is chosen at $10^{11}$ g/cm$^3$.
We pointed out that the frequency of the fundamental oscillations can be fitted by a function of 
the mass and radius of the PNS similar to the case of cold NSs.
In the case that the position of the outer boundary is chosen to cover not only the PNS but also the surrounding
postshock region, we obtained the eigenfrequencies close to the modulation frequencies of the SASI. On the other hand, our results suggested that these oscillation modes are unlikely to have the same physical origin of the SASI modes obtained in the hydrodynamics simulation. We have discussed possible limitations of applying the angle-averaged, linear perturbation analysis to extract the  full facets of the CCSN GW signatures. In order to identify the GW signatures in the spectrograms  more in a systematic manner, one may need to conduct a more detailed linear analysis as in Ref. \cite{TCOMF2019a}.

In this study we adopted the relativistic Cowling approximation, which could be applicable to the early postbounce phase because the stellar compactness is not so large and the relativistic effect may not be so significant. To apply the similar analysis to the late postbounce phase or to the very massive progenitor stars leading to a BH formation as reported in Ref.~\cite{Kuroda2018}, we need to perform the linear analysis taking into account the metric perturbation, which we shall leave for the future work. Towards the observation of the most remarkable spectral GW signature (i.e., 
the ramp-up $f$-mode) in the laser interferometers,  
dedicated data analysis schemes (e.g., \cite{astone2018}) need to be 
further developed.

\acknowledgments
This study was supported in part by the Grants-in-Aid for the Scientific Research of Japan Society for the Promotion of Science (JSPS, Nos. 
JP17K05458, 
JP26707013, 
JP17H01130, 
JP17K14306, 
JP18H01212  
), 
the Ministry of Education, Science and Culture of Japan (MEXT, Nos. 
JP15H00789, 
JP15H01039, 
JP15KK0173, 
JP17H05206 
JP17H06357, 
JP17H06364) 
 by the Central Research Institute of Fukuoka University (Nos.171042,177103) and the Research Institute of Explosive Stellar Phenomena (REISEP),
and JICFuS as a priority issue to be tackled by using Post `K' Computer.
TK was supported by the European Research Council (ERC; FP7) under ERC Starting Grant EUROPIUM-677912.



\end{document}